# Simplified modelling of chiral lattice materials with local resonators

Andrea Bacigalupo[1] and Luigi Gambarotta[2]

[1]IMT Institute for Advanced Studies Lucca, Italy
[2]Department of Civil, Chemical and Environmental Engineering, University of Genoa, Italy

**Abstract**

A simplified model of periodic chiral beam-lattices containing local resonators has been formulated to obtain a better understanding of the influence of the chirality and of the dynamic characteristics of the local resonators on the acoustic behavior. The simplified beam-lattices is made up of a periodic array of rigid heavy rings, each one connected to the others through elastic slender massless ligaments and containing an internal resonator made of a rigid disk in a soft elastic annulus. The band structure and the occurrence of low frequency band-gaps are analysed through a discrete Lagrangian model. For both the hexa- and the tetrachiral lattice, two acoustic modes and four optical modes are identified and the influence of the dynamic characteristics of the resonator on those branches is analyzed together with some properties of the band structure. By approximating the generalized displacements of the rings of the discrete Lagrangian model as a continuum field and through an application of the generalized macro-homogeneity condition, a generalized micropolar equivalent continuum has been derived, together with the overall equation of motion and the constitutive equation given in closed form. The validity limits of the micropolar model with respect to the dispersion functions are assessed by comparing the dispersion curves of this model in the irreducible Brillouin domain with those obtained by the discrete model, which are exact within the assumptions of the proposed simplified model.

# 1. Introduction

It is well known that the propagation of elastic waves may be strongly affected by periodic arrangement of scatterers in the material microstructure, in which the density and the elastic constants are periodic function of the position. This has spurred many researches on new materials such as phononic crystals and metamaterials for the control of vibrational waves (see Lu *et al.*, 2009, Pennec *et al.* 2010, Deymier, 2013, and Craster and Guenneau, 2013). In fact, the periodicity of the material microstructure may lead to destructive interferences inducing attenuation of the amplitude of the travelling waves for some bands of frequencies called acoustic wave spectral gap or band gaps.

Lattice materials are phononic crystals whose properties have been extensively studied from the seminal book of Brillouin (1953). Phani *et al.*, 2006, studied the dispersive wave propagation in periodic beam-lattices and showed the presence of band gaps. The analyzed model was made up of elastic slender beams rigidly connected at the nodes undergoing to axial strain, shearing and bending and subjected to inertial forces associated with the distributed mass of the beams themselves. Of special interest in this class of materials is the influence of the lattice topology on the acoustic properties (see Wang *et al.*, 2015). Auxetic lattices (see Prawoto, 2012) have attracted particular interest because of their dispersive properties (see Krödel *et al.*, 2014). Some theoretical and experimental studies on the phononic properties were carried out on chiral auxetic lattices, introduced in a seminal paper by Lakes, 1987. Particular attention has been devoted to hexachiral lattices, made up of circular rings each of them connected to its neighbors with six ligaments tangent to the ring itself, whose constitutive equation were firstly obtained by Prall and Lakes, 1997. Spadoni *et al.*, 2009, numerically investigated the dispersive waves in hexachiral lattice made up of elastic rings and ligaments with distributed mass. The periodic cell was analyzed with Bloch boundary conditions for several ratios between the length of the ligaments and the diameter of the rings and band gaps in the frequency spectrum were obtained. A band gap structure for plane tetrachiral lattices was experimentally obtained and numerically simulated by Tee *et al.*, 2010. These chiral beam-lattices have also described through equivalent continua mainly based on the micropolar model (see Spadoni and Ruzzene, 2012, Liu *et al.*, 2012, Chen *et al.*, 2014, Bacigalupo and Gambarotta, 2014a, Bacigalupo and De Bellis, 2015). Dispersive functions for hexachiral and tetrachiral lattices have been obtained by Liu *et al.*, 2012, and Chen *et al,*, 2014, respectively. However, in the domain of considered wave numbers, these equivalent micropolar models do not show band gaps.

To obtain low frequency band gaps, the insertion in the microstructure of local resonators generally made of a hand core surrounded by a soft coating has been proved effective. In fact, the locally resonant material may exhibit the emergence of stop bands at frequencies around the natural frequency of the resonator with overall negative mass density and bulk modulus (see for instance Liu *et al.*, 2000, Huang *et al.*, 2009a, b, Lai *et al.*, 2011, Raghavan and Srikantha Phani, 2013, Krushynska *et al.*, 2014). Chiral periodic metamaterials with internal locally resonant structures supporting tunable low-frequency stop bands have been recently proposed by Liu *et al.*, 2011a, Bigoni *et al.*, 2013, and Zhu *et al.*, 2014. In particular, Liu *et al.*, 2011a, have shown the benefice resulting from the chiral microstructure with local resonators that allow coupling the local translational and rotational resonances. Hexachiral beam-lattices integrated with local resonators made up of a softly coated heavy cylinder located inside the rings were analysed numerically by Liu *et al.*, 2011b, for low-frequency wave applications. Through a finite element analysis of the periodic cell with Bloch boundary conditions, the dispersive functions were derived in the reduced Brillouin domain and low-frequency band gaps were obtained.

The present paper is focused on understanding of the acoustic behavior of these chiral beam-lattice models with reference to different aims. A first issue concerns the sensitivity of the acoustic behavior and the formation of low-frequency band gaps as a result of the chiral geometry and the dynamic characteristics of the local resonators. Although some studies on the optimization of band gap in acoustic metamaterials have been carried out (see Tan *et al.*, 2012), in this study is preferred to consider a simple model of hexachiral and tetrachiral lattices (see Figure 1) having a reduced number of degrees of freedom and therefore able to provide some useful analytical results. To this end, the rings are assumed to be rigid and equipped with mass, as well as the cylindrical mass of the resonator, while the ligaments are assumed elastic but massless. These ones are connected to the rings according to various configurations, ranging from the achiral geometry, with the ligaments normal to the ring, to that of maximum chirality, with the ligaments tangent to the ring. These assumptions appear realistic, because the rings can be manufactured with these properties and the vibrations of the ligaments occur at high frequencies (see Phani *et al.*, 2006), while the interest of this study is more aimed at low frequencies. A simple Lagrangian model is formulated, which allows the determination of dispersive elastic waves and provides a simple evaluation and a comparison of the effects of the chirality with those of the local resonators.

A further issue concerns the formulation of a homogenized continuum model equivalent to the discrete Lagrangian. The homogenization of beam-lattice models has been tackled by several authors generally referring to homogeneous micropolar models (see for reference Bazant and Christensen, 1972, Noor *et al.*, 1978, Chen *et al.*, 1998, Pradel and Sab, 1998, Forest and Pradel , 2001, Onk, 2002, Ostoja-Starzewski, 2002, Kumar and McDowell, 2004, Gonnella and Ruzzene, 2008a,b, Bacigalupo and Gambarotta 2014a,b). On the other side, the wave propagation analysis through the dynamic homogenization of beam lattices has been analyzed by Suiker *et al.*, 2001, Ostoja-Starzewski, 2002, Gonnella and Ruzzene, 2008b, Stefanou *et al.*, 2008, Vasiliev *et al.*, 2010. The discrete model above described is here homogenized through a generalized energy equivalence criterion, by considering an approximation of the generalized displacement field through a second order Taylor expansion and by applying an appropriate transformation already proposed by Bazant and Christensen, 1972, and then taken up by Kumar and McDowell, 2004 and Liu *et al.*, 2012. The equations of motion thus obtained are those of a generalized micropolar continuum characterized by a generalized displacement field equipped with six degrees of freedom. It may be shown that these equations coincide with those derived by substituting the second order Taylor approximation of the displacement field in the equation of motion of the discrete model.

In order to investigate the influence of chirality on low frequency band gaps, both hexachiral and tetrachiral beam lattices are analysed, respectively, and the dispersion functions of the discrete and of the homogenized model are given, respectively, for several chiral angles measuring the inclination of the ligaments with respect to the line grid joining the centres of the rings. For both the lattices, two acoustic modes and four optical modes are identified and the influence of the dynamic characteristics of the resonator on those branches is analyzed together with some properties of the band structure. The validity limits of the micropolar model with respect to the dispersion functions are assessed by comparing the dispersion curves of this model in the irreducible Brillouin domain with those obtained by the discrete model, which are exact within the assumptions of the proposed simplified model.

## 2. Chiral lattice with local resonators: a simplified model

The beam-lattices shown in Figure 1 are based on the hexachiral and tetrachiral periodic cells shown in Figure 2, respectively. Each cell having size $a$ is made up of a ring with mean radius $r$ and $n$ ($=4,6$) slender ligaments of length $l$, section width $t$ and unit thickness, rigidly connected to the rings. The inclination of each ligament is denoted by the angle $\beta$ with respect to the lines connecting the centres of the rings. A heavy disk with external radius $R$ shown in Figure 2 (in dark grey), is located inside the ring through a soft elastic annulus (in yellow). This inclusion plays the role of low-frequency resonator. Increasing the angle $\beta$, a chiral microstructure with auxetic behaviour is obtained up to the condition in which the ligaments are tangent to the ring, when the angle takes the value $\beta_m = \arcsin\left(\dfrac{2r}{a}\right)$. This geometry allows to consider separately the effects of both the chiral microstructure (increasing $\beta$) and of the local resonator (increasing $R$) on the acoustic behaviour of the beam lattice. For $\beta \to 0$ the microstructure is no longer chiral, while for $R \to 0$ the resonator disappears. It follows that the independent geometric parameters of the model are: $a$, $r$, $R$, $t$ and $\beta$. The hexachiral lattice is transversely isotropic while the tetrachiral material belongs to the tetragonal system (see Bacigalupo and Gambarotta, 2014a,b).

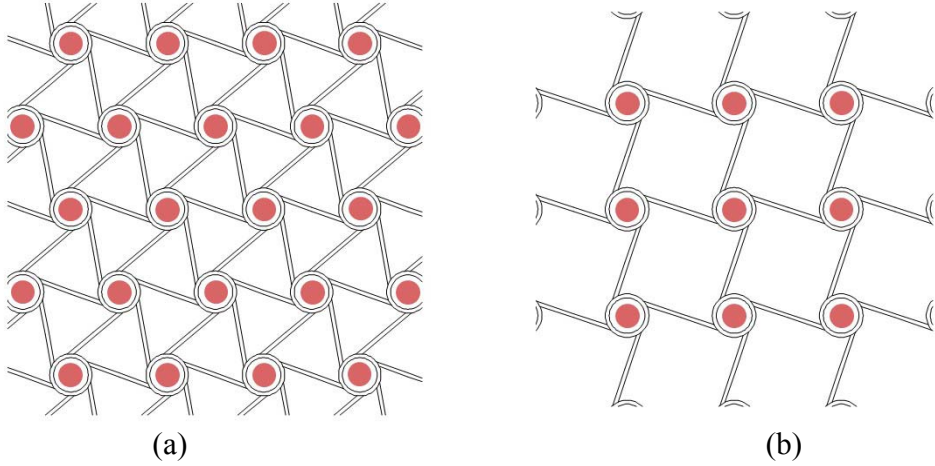

(a) (b)

Figure 1: (a) Hexachiral lattice; (b) tetrachiral lattice.

To obtain a simplified dynamic model (compare with Liu *et al.*, 2011) the rings and the disc of the resonators are assumed rigid, a technological condition that can be easily accomplished, while the inertia of the elastic soft coating and of the ligaments are ignored. From this last hypothesis, the high frequency vibrations of the ligaments are ignored, an aspect that is assumed not relevant to the present study which is focused on low frequency band gaps. The Young modulus of the ligaments is denoted by $E_s$, while the rings have mass density $\rho_s$, so that the translational and the rotatory inertia of the rings are $M_1 = 2\pi\rho_s rt$ and $J_1 = M_1 r^2$, respectively. The soft elastic coating inside the resonator has Young's modulus $E_a$ and Poisson's ratio $\nu_a$. The mass density of the internal resonator is denoted by $\rho_a$, so that its translational and the rotatory inertia are $M_2 = \pi\rho_a R^2$ and $J_2 = \frac{1}{2}M_2 R^2$, respectively.

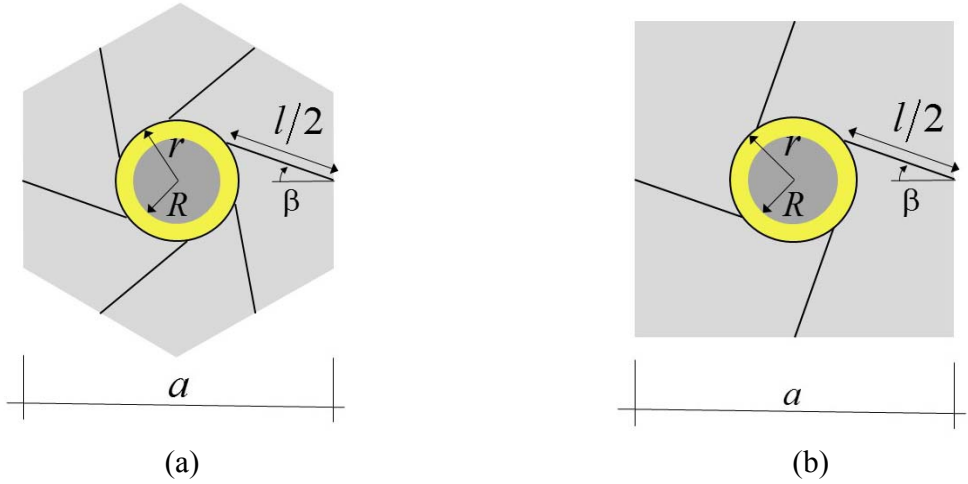

Figure 2: Periodic cell of the (a) hexachiral lattice; (b) tetrachiral lattice.

The motion of the rigid ring of the beam lattice is denoted by the displacement vector **u** and the rotation $\phi$, respectively (see Figure 3.a), while the motion of the internal resonator is denoted by the displacement vector **v** and the rotation $\theta$ (see Figure 3.b).

The constitutive equation of the soft elastic annulus connecting the internal rigid mass to the external rigid ring is

$$\mathbf{f} = -k_d(\mathbf{v}-\mathbf{u}) \quad , \quad c = -k_\theta(\theta-\phi) \quad , \tag{1}$$

**f** being the force exerted by the rigid ring on the internal mass and c the corresponding couple (see Figure 4). The translational and rotational stiffness $k_d$ and $k_\theta$, respectively, depend on the isotropic elastic moduli of the soft coating ($E_a, \nu_a$) and on the inner and outer radii as detailed in Appendix A.

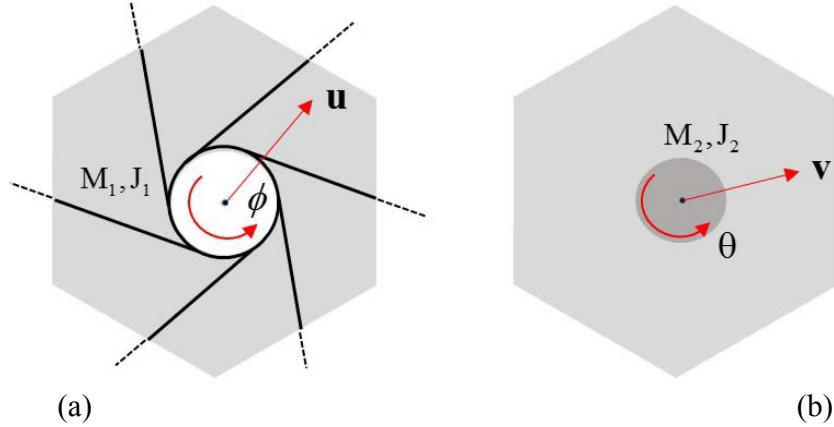

Figure 3: (a) The rigid ring and related dofs; (b) internal mass and related dofs.

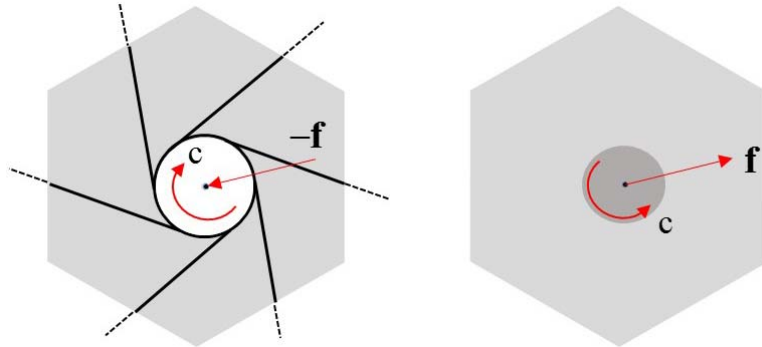

Figure 4: Contact force and couple between the rigid ring and the resonator.

The analysis of the propagation of elastic waves in two-dimensional lattices is carried out considering two models. The first, which is rigorous within the simplifying assumptions adopted, is based on a Lagrangian description characterized by the degrees of freedom of each ring together with those of the internal resonator. The second is based on an approximate description of the equations of motion through a dynamic homogenization in which the motion is described at the macroscale.

## 3. Dispersive wave propagation: the Lagrangian model

In the Lagrangian model the dissipative processes of deformation are ignored and six degrees of freedom for each cell are considered (three dofs for the ring, three dofs for the resonator). The equations of motion of the discrete dynamical system are obtained via the Lagrangian function $\mathcal{L} = \mathrm{T} - \Pi$, T being the kinetic energy and $\Pi$ the total potential elastic energy, sum of the contributions associated to all the elementary cells each centered in the corresponding ring. The contributions to the kinetic energy of the ring and the resonator inside the reference unit cell are

$$T_s\left(\dot{\mathbf{u}}, \dot{\phi}\right) = \frac{1}{2} M_1 \dot{\mathbf{u}} \cdot \dot{\mathbf{u}} + \frac{1}{2} J_1 \dot{\phi}^2,$$
$$T_r\left(\dot{\mathbf{v}}, \dot{\theta}\right) = \frac{1}{2} M_2 \dot{\mathbf{v}} \cdot \dot{\mathbf{v}} + \frac{1}{2} J_2 \dot{\theta}^2. \quad (2)$$

The elastic potential energy in the reference cell is obtained as the superposition of the contributions due to the elastic energy stored in the soft annulus of the resonator and in the *n* ligaments surrounding the ring. The first contribution is

$$\Pi_r\left(\mathbf{u}, \phi, \mathbf{v}, \theta\right) = \frac{1}{2} k_d \left(\mathbf{v} - \mathbf{u}\right) \cdot \left(\mathbf{v} - \mathbf{u}\right) + \frac{1}{2} k_\theta \left(\theta - \phi\right)^2 \quad (3)$$

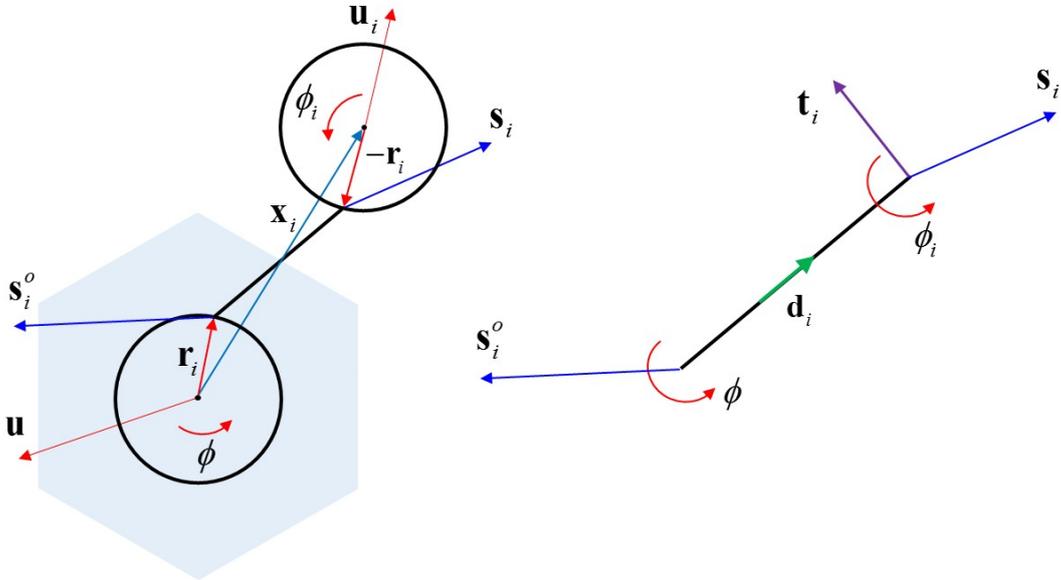

Figure 5: The *i*-th ligament between two adjacent cells: cell dofs and beam end displacements.

The elastic potential energy in the *i*-th ligament connecting the central reference ring of the reference cell to the *i*-th adjacent ring (see figure 5) depends on the end displacements of the ligament itself

$$\mathbf{s}_i^0 = \mathbf{u} + \phi\, \mathbf{e}_3 \times \mathbf{r}_i \quad , \quad \mathbf{s}_i = \mathbf{u}_i - \phi_i\, \mathbf{e}_3 \times \mathbf{r}_i \tag{4}$$

with $\mathbf{u}$, $\phi$, $\mathbf{u}_i$, $\phi_i$ the displacement rotation of the central ring and of the *i*-th ring, respectively, and vector $\mathbf{r}_i$ $(-\mathbf{r}_i)$ connecting the centre of the ring with the point of connection with the ligament as shown in Figure 5 and $\mathbf{e}_3$ the unit vector normal to the lattice. Here, the centre of the coordinated is assumed located in the centre of the central reference ring. Accordingly, the position of the *i*-th ring is $\mathbf{x}_i = a\, \mathbf{n}_i$, being $\mathbf{n}_i$ a unit vector. Moreover, the unit vector $\mathbf{d}_i$ represents the ligament orientation and $\mathbf{t}_i = \mathbf{e}_3 \times \mathbf{d}_i$ is the unit vector normal to $\mathbf{d}_i$. Note that the relation holds $\mathbf{r}_i = \dfrac{a}{2}\mathbf{n}_i - \dfrac{l}{2}\mathbf{d}_i$. The ligament extension is

$$\Delta_{di} = (\mathbf{s}_i - \mathbf{s}_i^0)\cdot \mathbf{d}_i = (\mathbf{u}_i - \mathbf{u})\cdot \mathbf{d}_i - (\phi_i + \phi)(\mathbf{e}_3 \times \mathbf{r}_i)\cdot \mathbf{d}_i = (\mathbf{u}_i - \mathbf{u})\cdot \mathbf{d}_i + \frac{a}{2}\sin\beta\,(\phi_i + \phi). \tag{5}$$

The transverse relative displacement between the ends of the ligament is

$$\Delta_{ti} = (\mathbf{s}_i - \mathbf{s}_i^0)\cdot \mathbf{t}_i = (\mathbf{u}_i - \mathbf{u})\cdot \mathbf{t}_i - (\phi_i + \phi)(\mathbf{e}_3 \times \mathbf{r}_i)\cdot \mathbf{t}_i = (\mathbf{u}_i - \mathbf{u})\cdot \mathbf{t}_i - \frac{1}{2}(a\cos\beta - l)(\phi_i + \phi). \tag{6}$$

The mean rotation of the *i*-th ligament is $\tilde{\psi}_i = \Delta_{ti}/l$ and the end rotation of the ligament at the central and *i*-th ring are $\varphi_i^0 = \phi - \tilde{\psi}_i = \phi - \dfrac{\Delta_{ti}}{l}$ and $\varphi_i = \phi_i - \tilde{\psi}_i = \phi_i - \dfrac{\Delta_{ti}}{l}$, respectively.

The axial potential elastic energy due to axial strain $\varepsilon_i = \dfrac{\Delta_{di}}{l}$ of the *i*-th ligaments takes the form

$$\Pi_{ai}(\mathbf{u}, \phi, \mathbf{u}_i, \phi_i) = \frac{1}{2} E_s t \frac{\Delta_{di}^2}{l} \tag{7}$$

and the bending potential elastic energy due to bending of the ligament is written as

$$\Pi_{bi}\left(\mathbf{u},\phi,\mathbf{u}_i,\phi_i\right)=\frac{E_s t^2}{6}\left(\frac{t}{l}\right)\left(\varphi_i^{0\,2}+\varphi_i^2+\varphi_i^0\,\varphi_i\right). \tag{8}$$

The resulting Lagrangian function is obtained as the sum of the contributions of all the elementary cells

$$\mathcal{L}=\sum\left\{T_s\left(\dot{\mathbf{u}},\dot{\phi}\right)+T_r\left(\dot{\mathbf{v}},\dot{\theta}\right)-\Pi_r\left(\mathbf{u},\phi,\mathbf{v},\theta\right)-\sum_{i=1}^{n}\left[\Pi_{ai}\left(\mathbf{u},\phi,\mathbf{u}_i,\phi_i\right)+\Pi_{bi}\left(\mathbf{u},\phi,\mathbf{u}_i,\phi_i\right)\right]\right\} \tag{9}$$

where the index denoting each cell has been omitted for simplicity and each contribution is obtained according to the above evaluation. The Euler-Lagrange equations of motion of the considered cell are obtained by the Lagrangian function (9) and depend on the generalized displacement and velocity of the ring and the resonator at the center of the cell and on the generalized displacement and velocity of the rings of the surrounding cell. The equation of motion of the reference cell are written as a system of six ODEs taking the form

$$E_s\left(\frac{t}{l}\right)\sum_{i=1}^{n}\left[\tilde{\mathbf{A}}_i\left(\mathbf{u}_i-\mathbf{u}\right)-\frac{a}{2}\tilde{\mathbf{b}}_i\left(\phi+\phi_i\right)\right]+k_d\left(\mathbf{v}-\mathbf{u}\right)-M_1\ddot{\mathbf{u}}=\mathbf{0}, \tag{10}$$

$$E_s\left(\frac{t}{l}\right)\sum_{i=1}^{n}\left\{\begin{array}{c}\frac{a}{2}\tilde{\mathbf{b}}_i\cdot\left(\mathbf{u}_i-\mathbf{u}\right)+\\-\frac{a^2}{4}\left[\sin^2\beta+\left(\frac{t}{l}\right)^2\cos^2\beta\right]\left(\phi+\phi_i\right)+\frac{l^2}{12}\left(\frac{t}{l}\right)^2\left(\phi_i-\phi\right)\end{array}\right\}+k_\theta\left(\theta-\phi\right)-J_1\ddot{\phi}=0, \tag{11}$$

$$k_d\left(\mathbf{v}-\mathbf{u}\right)+M_2\ddot{\mathbf{v}}=\mathbf{0}, \tag{12}$$

$$k_\theta\left(\theta-\phi\right)+J_2\ddot{\theta}=0, \tag{13}$$

having defined $\tilde{\mathbf{A}}_i=\left(\mathbf{d}_i\otimes\mathbf{d}_i\right)+\left(\frac{t}{l}\right)^2\left(\mathbf{t}_i\otimes\mathbf{t}_i\right)$ and $\tilde{\mathbf{b}}_i=-\sin\beta\,\mathbf{d}_i+\left(\frac{t}{l}\right)^2\cos\beta\,\mathbf{t}_i$.

If a harmonic plane wave propagating along axis $\mathbf{i}$ in an infinite two-dimensional medium is admitted, the generalized displacement field at a point is assumed in the following form $\mathbf{U}=\hat{\mathbf{U}}\exp\left[i\left(\mathbf{k}\cdot\mathbf{x}-\omega t\right)\right]$, where $\mathbf{k}=q\mathbf{i}$ is the wave vector and $q$ and $\omega$ denote the wave number and the circular frequency, respectively, and $\hat{\mathbf{U}}=\left\{\hat{\mathbf{u}}^T\ \hat{\phi}\ \hat{\mathbf{v}}^T\ \hat{\theta}\right\}^T=\left\{\hat{u}_1\ \hat{u}_2\ \hat{\phi}\ \hat{v}_1\ \hat{v}_2\ \hat{\theta}\right\}^T$ is the polarization vector. Substituting the assumed generalized displacement field in the equation of motion (10) ÷ (13) and noting that

$$\mathbf{u}_i - \mathbf{u} = \hat{\mathbf{u}}\left[\exp(i\mathbf{k}\cdot\mathbf{x}_i)-1\right]\exp(-i\omega t), \qquad \phi_i - \phi = \hat{\phi}\left[\exp(i\mathbf{k}\cdot\mathbf{x}_i)-1\right]\exp(-i\omega t) \quad \text{and}$$

$\phi + \phi_i = \hat{\phi}\left[1+\exp(i\mathbf{k}\cdot\mathbf{x}_i)\right]\exp(-i\omega t)$, one obtains the system of six linear homogeneous equations:

$$\mathbf{C}_{Lag}(\mathbf{k},\omega)\hat{\mathbf{U}} =$$

$$= \begin{bmatrix} E_s\left(\dfrac{t}{l}\right)\mathbf{A}+k_d\mathbf{I}_2-\omega^2 M_1\mathbf{I}_2 & E_s\left(\dfrac{t}{l}\right)\mathbf{b}^+ & -k_d\mathbf{I}_2 & \mathbf{0}_{2\times1} \\ E_s\left(\dfrac{t}{l}\right)\mathbf{b}^- & E_s\left(\dfrac{t}{l}\right)C+k_\theta-\omega^2 J_1 & \mathbf{0}_{1\times2} & -k_\theta \\ -k_d\mathbf{I}_2 & \mathbf{0}_{2\times1} & k_d\mathbf{I}_2-\omega^2 M_2\mathbf{I}_2 & \mathbf{0}_{2\times1} \\ \mathbf{0}_{1\times2} & -k_\theta & \mathbf{0}_{1\times2} & k_\theta-\omega^2 J_2 \end{bmatrix} \begin{Bmatrix} \hat{\mathbf{u}} \\ \hat{\phi} \\ \hat{\mathbf{v}} \\ \hat{\theta} \end{Bmatrix} = \mathbf{0} \quad (14)$$

where the following terms are defined in view of the centrosymmetric layout of the ligaments in the periodic cell:

$$\mathbf{A} = \sum_{i=1}^{n}\left[1-\exp(i\mathbf{k}\cdot\mathbf{x}_i)\right]\tilde{\mathbf{A}}_i = \sum_{i=1}^{n}\left[1-\cos(\mathbf{k}\cdot\mathbf{x}_i)\right]\tilde{\mathbf{A}}_i, \qquad (15)$$

$$\mathbf{b}^+ = \frac{a}{2}\sum_{i=1}^{n}\left[1+\exp(i\mathbf{k}\cdot\mathbf{x}_i)\right]\tilde{\mathbf{b}}_i = i\,\frac{a}{2}\sum_{i=1}^{n}\sin(\mathbf{k}\cdot\mathbf{x}_i)\tilde{\mathbf{b}}_i, \qquad (16)$$

$$\mathbf{b}^- = \frac{a}{2}\sum_{i=1}^{n}\left[1-\exp(i\mathbf{k}\cdot\mathbf{x}_i)\right]\tilde{\mathbf{b}}_i^T = -i\,\frac{a}{2}\sum_{i=1}^{n}\sin(\mathbf{k}\cdot\mathbf{x}_i)\tilde{\mathbf{b}}_i^T, \qquad (17)$$

$$C = \sum_{i=1}^{n}\left\{\frac{a^2}{4}\left[\sin^2\beta+\left(\frac{t}{l}\right)^2\cos^2\beta\right]\left[1+\exp(i\mathbf{k}\cdot\mathbf{x}_i)\right]+\frac{l^2}{12}\left(\frac{t}{l}\right)^2\left[1-\exp(i\mathbf{k}\cdot\mathbf{x}_i)\right]\right\} =$$

$$= \sum_{i=1}^{n}\left\{\frac{a^2}{4}\left[\sin^2\beta+\left(\frac{t}{l}\right)^2\cos^2\beta\right]\left[1+\cos(\mathbf{k}\cdot\mathbf{x}_i)\right]+\frac{l^2}{12}\left(\frac{t}{l}\right)^2\left[1-\cos(\mathbf{k}\cdot\mathbf{x}_i)\right]\right\} \qquad (18)$$

with $\mathbf{I}_2$ unit matrix and $\mathbf{0}_{1\times2}$ zero vector of order 2, respectively. Because of the centrosymmetry of the cell it follows $\text{Im}[\mathbf{A}]=\mathbf{0}$, $\text{Im}[C]=0$, $\text{Re}[\mathbf{b}^+]=\text{Re}[\mathbf{b}^-]=\mathbf{0}$ and $\text{Im}[\mathbf{b}^+]=-\text{Im}\left[(\mathbf{b}^-)^T\right]$ and the matrix $\mathbf{C}_{Lag}(\mathbf{k},\omega)$ is Hermitian.

The angular frequency $\omega(\mathbf{k})$ and the polarization vector $\hat{\mathbf{U}}(\mathbf{k})$ of a travelling wave with wave vector $\mathbf{k}$ is obtained by solving the eigenvalue (14), from which six dispersive branches are obtained in the irreducible Brillouin zone. In the long wavelength limit, namely for $|\mathbf{k}| \to 0$, one obtains $\mathbf{A} = \mathbf{0}$, $\mathbf{b}^+ = \mathbf{b}^- = \mathbf{0}$ and $C_0 = C(|\mathbf{k}| \to 0) = \dfrac{na^2}{2}\left[\sin^2\beta + \left(\dfrac{t}{l}\right)^2 \cos^2\beta\right]$, and a double vanishing solution is obtained from which two acoustic branches take place. Another double solution is obtained $\omega_{opt3,4} = \sqrt{1 + \dfrac{M_2}{M_1}}\,\omega_d$, being $\omega_d = \sqrt{k_d/M_2}$ the translational frequency of the resonator, that defines a critical point with vanishing group velocity ($v_g = \left.\dfrac{d\omega(\mathbf{k})}{d|\mathbf{k}|}\right|_{|\mathbf{k}|\to 0} = 0$) on the band structure from which two optical branches depart. Finally, two different solutions are obtained $\omega_{opt5,6} = \dfrac{\sqrt{2}}{2}\sqrt{1 + \dfrac{J_2}{J_1} + \left(\dfrac{\omega_\phi}{\omega_\theta}\right)^2 \mp \sqrt{\left[1 + \dfrac{J_2}{J_1} + \left(\dfrac{\omega_\phi}{\omega_\theta}\right)^2\right]^2 - 4\left(\dfrac{\omega_\phi}{\omega_\theta}\right)^2}}\;\omega_\theta$, being $\omega_\theta = \sqrt{k_\theta/J_2}$ the rotational frequency of the resonator and $\omega_\phi = \sqrt{H/J_1}$ ($H = E_s\left(\dfrac{t}{l}\right)C_0$) the frequency of the lattice with no resonators at the long wavelength limit ($|\mathbf{k}| \to 0$). These two critical points define in the band structure two corresponding optical branches.

## 4. Equation of motion in the equivalent continuum model

An approximation to the description of motion of the Lagrangian system considered in the previous paragraph can be obtained by introducing continuous fields of displacement and rotation to describe the generalized displacement of the rings and of resonators. The displacement vector and the rotation of the ring of the *i*-th neighbouring cell may be approximated through a second-order Taylor expansion according to Bazant and Christensen (1972) and Kumar and McDowell (2004)

$$\begin{aligned}\mathbf{u}_i &\simeq \mathbf{u} + \mathbf{H}\mathbf{x}_i + \frac{1}{2}\nabla\mathbf{H}:(\mathbf{x}_i \otimes \mathbf{x}_i),\\ \phi_i &\simeq \phi + \boldsymbol{\chi}\cdot\mathbf{x}_i + \frac{1}{2}\nabla\boldsymbol{\chi}:(\mathbf{x}_i \otimes \mathbf{x}_i),\end{aligned} \qquad (19)$$

$\mathbf{H} = \nabla\mathbf{u}$ and $\nabla\mathbf{H}$ being the macro-displacement gradient and second gradient, $\mathbf{\chi} = \nabla\phi$ and $\nabla\mathbf{\chi}$ the curvature and its gradient tensor, respectively. In this case, being $\mathbf{x}_i = a\mathbf{n}_i$, the extension of the $i$-th ligament takes the form

$$\Delta_{di} \simeq a(\mathbf{d}_i \otimes \mathbf{n}_i):\mathbf{\Gamma} + \frac{a^2}{2}\sin\beta\, \mathbf{n}_i \cdot \mathbf{\chi} + \frac{a^2}{2}\nabla\mathbf{H}\vdots(\mathbf{d}_i \otimes \mathbf{n}_i \otimes \mathbf{n}_i) + \frac{a^3}{4}\sin\beta\, \nabla\mathbf{\chi}:(\mathbf{n}_i \otimes \mathbf{n}_i), \quad (20)$$

where the Cosserat strain tensor $\mathbf{\Gamma} = \mathbf{H} - \mathbf{W}(\phi)$ and the rotation tensor $\mathbf{W}(\phi) = \begin{bmatrix} 0 & -\phi \\ \phi & 0 \end{bmatrix}$ are introduced (note that $(\mathbf{e}_3 \times \mathbf{n}_i) \cdot \mathbf{d}_i = -\sin\beta$ and $\phi\, \mathbf{e}_3 \times \mathbf{n}_i = \mathbf{W}(\phi)\mathbf{n}_i$). Moreover, the transversal displacement is obtained as

$$\Delta_{ti} \simeq \phi l + a(\mathbf{t}_i \otimes \mathbf{n}_i):\mathbf{\Gamma} + \frac{a^2}{2}\nabla\mathbf{H}\vdots(\mathbf{t}_i \otimes \mathbf{n}_i \otimes \mathbf{n}_i) + \\
- \frac{a}{2}(a\cos\beta - l)\mathbf{\chi} \cdot \mathbf{n}_i - \frac{a^2}{4}(a\cos\beta - l)\left[\nabla\mathbf{\chi}:(\mathbf{n}_i \otimes \mathbf{n}_i)\right]. \quad (21)$$

The effective end rotation of the ligaments, being $\phi = \mathbf{t}_i \cdot \mathbf{W}\mathbf{n}_i$, are

$$\varphi_i^0 = \phi - \frac{\Delta_{ti}}{l} = -\frac{a}{l}(\mathbf{t}_i \otimes \mathbf{n}_i):\mathbf{\Gamma} - \frac{a^2}{2l}\nabla\mathbf{H}\vdots(\mathbf{t}_i \otimes \mathbf{n}_i \otimes \mathbf{n}_i) + \\
+ \frac{a}{2}\left(\frac{a}{l}\cos\beta - 1\right)\mathbf{\chi} \cdot \mathbf{n}_i + \frac{a^2}{4}\left(\frac{a}{l}\cos\beta - 1\right)\left[\nabla\mathbf{\chi}:(\mathbf{n}_i \otimes \mathbf{n}_i)\right], \\
\varphi_i = \phi_i - \frac{\Delta_{ti}}{l} = -\frac{a}{l}(\mathbf{t}_i \otimes \mathbf{n}_i):\mathbf{\Gamma} - \frac{a^2}{2l}\nabla\mathbf{H}\vdots(\mathbf{t}_i \otimes \mathbf{n}_i \otimes \mathbf{n}_i) + \\
+ \frac{a}{2}\left(\frac{a}{l}\cos\beta + 1\right)\mathbf{\chi} \cdot \mathbf{n}_i + \frac{a^2}{4}\left(\frac{a}{l}\cos\beta + 1\right)\left[\nabla\mathbf{\chi}:(\mathbf{n}_i \otimes \mathbf{n}_i)\right]. \quad (22)$$

The axial elastic potential energy density is derived from definition (7) in the form $\pi_{ai} = \Pi_{ai}/2A_{cell}$. Under assumption (19) on the generalized displacements and in consideration of the centrosymmetry of the cell the elastic potential density takes the form

$$\pi_a = \frac{E_s a^2}{4A_{cell}}\left(\frac{t}{l}\right)\sum_{i=1}^{n}\left\{ \begin{array}{l} (\mathbf{t}_i \otimes \mathbf{n}_i):\mathbf{\Gamma} + \dfrac{a}{2}\sin\beta\, \mathbf{n}_i \cdot \mathbf{\chi} + \\ + \dfrac{a}{2}\nabla\mathbf{H}\vdots(\mathbf{d}_i \otimes \mathbf{n}_i \otimes \mathbf{n}_i) + \dfrac{a^2}{4}\sin\beta\, \nabla\mathbf{\chi}:(\mathbf{n}_i \otimes \mathbf{n}_i) \end{array} \right\}^2. \quad (23)$$

By expanding (23) one obtains some tensors of odd rank as dyadic product involving the vectors $\mathbf{n}_i$, $\mathbf{t}_i$ and $\mathbf{d}_i$ which are vanishing as a consequence of the centrosymmetry of the periodic cell. Moreover, according to Bazant and Christensen (1972) and Kumar and McDowell (2004), the terms depending on higher order of generalized displacements $\nabla \mathbf{H}$ and $\nabla \boldsymbol{\chi}$ are disregarded apart from the term involving the tensors $\mathbf{W}$ and $\nabla \boldsymbol{\chi}$, namely the term $\frac{a^2}{4}\sin\beta\left[(\mathbf{t}_i \otimes \mathbf{n}_i):\mathbf{W}\right]\left[\nabla\boldsymbol{\chi}:(\mathbf{n}_i \otimes \mathbf{n}_i)\right]$. In fact, since $(\mathbf{t}_i \otimes \mathbf{n}_i):\mathbf{W} = \phi$, this term takes the form $\frac{a^2}{4}\sin\beta\,\phi\,\nabla\boldsymbol{\chi}:(\mathbf{n}_i \otimes \mathbf{n}_i) = \frac{a^2}{4}\sin\beta\,\phi\,\phi_{,pq}n^i_p n^i_q$. By noting that the integral over the periodic cell $\int_{A_{cell}}\phi\,\phi_{,pq}n^i_p n^i_q da = -\int_{A_{cell}}\phi_{,p}\,\phi_{,q}n^i_p n^i_q da + \int_{\partial A_{cell}}\phi_{,p}\,\phi\,n^i_p ds = -\int_{A_{cell}}\boldsymbol{\chi}\cdot(\mathbf{n}_i \otimes \mathbf{n}_i)\boldsymbol{\chi}\,da + \int_{\partial A_{cell}}\phi\,\boldsymbol{\chi}\cdot\mathbf{n}_i ds$ is a quadratic form of the Cosserat curvature plus a boundary term, it follows that, the axial potential energy density may be written as a quadratic form of the Cosserat strain measures:

$$\pi_a = \frac{E_s a^2}{4 A_{cell}}\left(\frac{t}{l}\right)\sum_{i=1}^{n}\left\{\boldsymbol{\Gamma}:(\mathbf{d}_i \otimes \mathbf{n}_i \otimes \mathbf{d}_i \otimes \mathbf{n}_i)\boldsymbol{\Gamma} + \boldsymbol{\chi}\cdot\left[-\frac{a^2}{4}\sin^2\beta\,(\mathbf{n}_i \otimes \mathbf{n}_i)\right]\boldsymbol{\chi}\right\}. \tag{24}$$

Similarly, the bending elastic potential energy density $\pi_{bi} = \Pi_{bi}/2A_{cell}$ is considered. From equations (6) and approximation (19) and proceeding in analogy to the case of axial strain as previously shown, the quadratic form is obtained

$$\pi_b = \frac{E_s a^2}{12 A_{cell}}\left(\frac{t}{l}\right)^3 \sum_{i=1}^{n}\left\{3\boldsymbol{\Gamma}:(\mathbf{t}_i \otimes \mathbf{n}_i \otimes \mathbf{t}_i \otimes \mathbf{n}_i)\boldsymbol{\Gamma} + \boldsymbol{\chi}\cdot\left[-\frac{a^2}{4}\left(3\cos^2\beta - \frac{l^2}{a^2}\right)(\mathbf{n}_i \otimes \mathbf{n}_i)\right]\boldsymbol{\chi}\right\}. \tag{25}$$

Lastly, the density of elastic potential energy is obtained as a quadratic form of a classical centrosymmetric micropolar continuum.

$$\pi_s = \frac{1}{2}\boldsymbol{\Gamma}\bullet\mathbb{E}_s\boldsymbol{\Gamma} + \frac{1}{2}\boldsymbol{\chi}\cdot\mathbf{E}_s\boldsymbol{\chi}, \tag{26}$$

where the fourth order elasticity tensor

$$\mathbb{E}_s = \frac{E_s a^2}{2 A_{cell}}\left(\frac{t}{l}\right)\sum_{i=1}^{n}\left[(\mathbf{d}_i \otimes \mathbf{n}_i \otimes \mathbf{d}_i \otimes \mathbf{n}_i) + \left(\frac{t}{l}\right)^2 (\mathbf{t}_i \otimes \mathbf{n}_i \otimes \mathbf{t}_i \otimes \mathbf{n}_i)\right], \tag{27}$$

is endowed with the major symmetry and is positive defined while the symmetric second order elasticity tensor associated with the curvature

$$\mathbf{E}_s = -\frac{E_s a^4}{24 A_{cell}}\left(\frac{t}{l}\right)\left\{3\sin^2\beta + \left(\frac{t}{l}\right)^2\left[3\cos^2\beta - \left(\frac{l}{a}\right)^2\right]\right\}\sum_{i=1}^n (\mathbf{n}_i \otimes \mathbf{n}_i) \ . \qquad (28)$$

is negative defined. Conversely, if the second order terms of the generalized displacement field of the rings are neglected in the approximation (19) (see for instance Chen *et al.*, 1998) the second order elasticity tensor takes the following positive defined form

$$\mathbf{E}_s^+ = \frac{E_s a^4}{24 A_{cell}}\left(\frac{t}{l}\right)\left\{3\sin^2\beta + \left(\frac{t}{l}\right)^2\left[3\cos^2\beta + \left(\frac{l}{a}\right)^2\right]\right\}\sum_{i=1}^n (\mathbf{n}_i \otimes \mathbf{n}_i) \ . \qquad (29)$$

The elastic potential energy density stored in the resonator is defined as follows

$$\pi_r = \frac{1}{2}\hat{k}_d (\mathbf{v}-\mathbf{u})\cdot(\mathbf{v}-\mathbf{u}) + \frac{1}{2}\hat{k}_\theta (\theta-\phi)^2 \ , \qquad (30)$$

$\hat{k}_d = k_d / A_{cell}$ and $\hat{k}_\theta = k_\theta / A_{cell}$ being the averaged stiffnesses of the resonator. The kinetic energy density of the ring and of the resonator takes, respectively, the forms

$$\begin{aligned} t_s &= \frac{1}{2}\rho_1 \dot{\mathbf{u}}\cdot\dot{\mathbf{u}} + \frac{1}{2}I_1 \dot{\phi}^2, \\ t_r &= \frac{1}{2}\rho_2 \dot{\mathbf{v}}\cdot\dot{\mathbf{v}} + \frac{1}{2}I_2 \dot{\theta}^2. \end{aligned} \qquad (31)$$

where the averaged mass densities $\rho_1 = M_1 / A_{cell}$ and $\rho_2 = M_2 / A_{cell}$, and the micro-inertia terms $I_1 = J_1 / A_{cell}$ and $I_2 = J_2 / A_{cell}$ of the ring and the resonator are defined, respectively.

The Lagrangian defined on the periodic cell takes the form

$$\mathcal{L} = \int_{A_{cell}} \left[ \begin{array}{c} \frac{1}{2}\rho_1 \dot{\mathbf{u}}\cdot\dot{\mathbf{u}} + \frac{1}{2}\mathrm{I}_1\dot{\phi}^2 + \frac{1}{2}\rho_2 \dot{\mathbf{v}}\cdot\dot{\mathbf{v}} + \frac{1}{2}\mathrm{I}_2\dot{\theta}^2 + \\ -\frac{1}{2}\boldsymbol{\Gamma}\bullet\mathbb{E}_s\boldsymbol{\Gamma} - \frac{1}{2}\boldsymbol{\chi}\cdot\mathbf{E}_s\boldsymbol{\chi} - \frac{1}{2}\hat{k}_d(\mathbf{v}-\mathbf{u})\cdot(\mathbf{v}-\mathbf{u}) - \frac{1}{2}\hat{k}_\theta(\theta-\phi)^2 \end{array} \right] da + bound.\ terms \quad (32)$$

and from the application of the Hamilton's extended principle, the equations of motion of an enhanced micropolar continuum are obtained in the form

$$\begin{cases} div(\mathbb{E}_s \boldsymbol{\Gamma}) + \hat{k}_d (\mathbf{v} - \mathbf{u}) = \rho_1 \ddot{\mathbf{u}}, \\ -\delta_{3jh} (\mathbf{e}_j \otimes \mathbf{e}_h) : \mathbb{E}_s \boldsymbol{\Gamma} + \mathbf{E}_s : \nabla \boldsymbol{\chi} + \hat{k}_\theta (\theta - \phi) = I_1 \ddot{\phi}, \\ \hat{k}_d (\mathbf{u} - \mathbf{v}) = \rho_2 \ddot{\mathbf{v}}, \\ \hat{k}_\theta (\phi - \theta) = I_2 \ddot{\theta}, \end{cases} \quad (33)$$

where $\delta_{ijk}$ is the Levi-Civita symbol. These equations can be obtained through an alternative procedure by substituting the approximation (19) of the generalized displacement of the *i*-th ring in the equations of motion of the discrete model (10) ÷ (13). It is worth to note that such circumstance is achieved if the negative defined elasticity tensor given in (28) is applied, but not with the positive definite tensor given in (29).

Finally, from the definition of elastic potential energy density (26) stored in the ligaments, the overall asymmetric stress tensor and the couple stress vector are obtained

$$\mathbf{T} = \frac{\partial \pi_s}{\partial \boldsymbol{\Gamma}} = \mathbb{E}_s \boldsymbol{\Gamma}, \qquad \mathbf{m} = \frac{\partial \pi_s}{\partial \boldsymbol{\chi}} = \mathbf{E}_s \boldsymbol{\chi} \quad (34)$$

having components $\sigma_{11}, \sigma_{12}, \sigma_{21}, \sigma_{22}$ and $m_1$ and $m_2$, respectively, which are energetically conjugated to the components $\gamma_{11} = u_{1,1}$, $\gamma_{22} = u_{2,2}$, $\gamma_{12} = u_{1,2} + \phi$, $\gamma_{21} = u_{2,1} - \phi$ of the overall asymmetric strain tensor and to the curvatures $\chi_1 = \phi_{,1}$ and $\chi_2 = \phi_{,2}$ of the chiral lattice.

## 5. Wave propagation in the equivalent generalized micropolar continuum

The equations of motion of the equivalent continuum derived in the previous Section are specialized to the cases of hexachiral and tetrachiral lattices. Hence, the equations governing the propagation of harmonic plane waves are formulated and the angular frequencies for the long-wavelength limit are derived.

*5.1 Hexachiral lattice*

For the hexachiral lattice, the elasticity tensors are derived from equations (27) and (28) noting that $A_{cell} = \sqrt{3} a^2 / 2$, and the constitutive equation is written in the Voigt notation

$$\begin{Bmatrix} \sigma_{11} \\ \sigma_{22} \\ \sigma_{12} \\ \sigma_{21} \\ m_1 \\ m_2 \end{Bmatrix} = \begin{bmatrix} 2\mu+\lambda & \lambda & -A & A & 0 & 0 \\ \lambda & 2\mu+\lambda & -A & A & 0 & 0 \\ -A & -A & \mu+\kappa & \mu-\kappa & 0 & 0 \\ A & A & \mu-\kappa & \mu+\kappa & 0 & 0 \\ 0 & 0 & 0 & 0 & S & 0 \\ 0 & 0 & 0 & 0 & 0 & S \end{bmatrix} \begin{Bmatrix} \gamma_{11} \\ \gamma_{22} \\ \gamma_{12} \\ \gamma_{21} \\ \chi_1 \\ \chi_2 \end{Bmatrix}, \quad (35)$$

with the five elastic moduli

$$\begin{aligned}
\mu &= \frac{\sqrt{3}}{4} E_s \left(\frac{t}{l}\right)\left[1+\left(\frac{t}{l}\right)^2\right], \\
\lambda &= \frac{\sqrt{3}}{4} E_s \left(\frac{t}{l}\right)\left[1-\left(\frac{t}{l}\right)^2\right]\cos 2\beta, \\
\kappa &= \frac{\sqrt{3}}{2} E_s \left(\frac{t}{l}\right)\left[\sin^2\beta + \left(\frac{t}{l}\right)^2 \cos^2\beta\right], \\
A &= -\frac{\sqrt{3}}{4} E_s \left(\frac{t}{l}\right)\left[1-\left(\frac{t}{l}\right)^2\right]\sin 2\beta, \\
S &= -\frac{\sqrt{3}}{12} E_s a^2 \left(\frac{t}{l}\right)\left\{3\left[\sin^2\beta + \left(\frac{t}{l}\right)^2 \cos^2\beta\right] - \left(\frac{t}{l}\right)^2\left(\frac{l}{a}\right)^2\right\}.
\end{aligned} \quad (36)$$

depending on the Young's modulus $E_s$, the characteristic size $a$ of the lattice, the slenderness ratio $t/l$ and the angle $\beta$ of inclination of the ligaments. It is worth to note that in case of ligaments tangent to the ring $(\beta_m)$ these equation corresponds to those obtained by Liu *et al.*, 2012, and Bacigalupo and Gambarotta, 2014. Moreover, if only a first order approximation of the generalized displacement field is assumed, the positive defined elastic constant is obtained $S^+ = \frac{\sqrt{3}}{12} E_s a^2 \left(\frac{t}{l}\right)\left\{3\left[\sin^2\beta + \left(\frac{t}{l}\right)^2 \cos^2\beta\right] + \left(\frac{t}{l}\right)^2\left(\frac{l}{a}\right)^2\right\}$. The elastic moduli, with the exception of parameter $\mu$, depend on the angle of chirality $\beta$, but only the constant $A$, that couples the extensional and the asymmetric strain components, is an odd function of this parameter. For symmetric macro-strain fields $\phi = \frac{1}{2}(u_{2,1} - u_{1,2})$ one obtains the in-plane elastic moduli of the transversely isotropic continuum

$$E_{\text{hom}} = \frac{2\sqrt{3} E_s \left(\frac{t}{l}\right)^3 \left[1 + \left(\frac{t}{l}\right)^2\right]}{\sin^2\beta + 3\left(\frac{t}{l}\right)^2 + \left(\frac{t}{l}\right)^4 \cos^2\beta} \quad,$$

$$\nu_{\text{hom}} = -\frac{\left[\sin^2\beta - \left(\frac{t}{l}\right)^2 \cos^2\beta\right]\left[1 - \left(\frac{t}{l}\right)^2\right]}{\sin^2\beta + 3\left(\frac{t}{l}\right)^2 + \left(\frac{t}{l}\right)^4 \cos^2\beta} \quad, \quad (37)$$

$$G_{\text{hom}} = \frac{\sqrt{3}}{4} E_s \left(\frac{t}{l}\right)^2 \left[1 + \left(\frac{t}{l}\right)^2\right] \quad.$$

The free wave motion is derived from equations (33) in terms of the components of the generalized displacement field $\mathbf{U}(\mathbf{x}) = \{u_1 \quad u_2 \quad \phi \quad v_1 \quad v_2 \quad \theta\}^T$ in the following form

$$\begin{cases}
(2\mu + \lambda)u_{1,11} - 2Au_{1,12} + (\mu + k)u_{1,22} + Au_{2,11} + (\mu - k + \lambda)u_{2,12} - Au_{2,22} + \\
\qquad\qquad\qquad - 2A\phi_{,1} + 2k\phi_{,2} + \hat{k}_d(v_1 - u_1) = \rho_1 \ddot{u}_1 \quad, \\
Au_{1,11} + (\mu - k + \lambda)u_{1,12} - Au_{1,22} + (\mu + k)u_{2,11} + 2Au_{2,12} + (2\mu + \lambda)u_{2,22} + \\
\qquad\qquad\qquad - 2k\phi_{,1} - 2A\phi_{,2} + \hat{k}_d(v_2 - u_2) = \rho_1 \ddot{u}_2 \quad, \\
S\phi_{,11} + S\phi_{,22} + 2Au_{1,1} + 2Au_{2,2} - 2ku_{1,2} + 2ku_{2,1} - 4k\phi + \hat{k}_\theta(\theta - \phi) = I_1 \ddot{\phi} \quad, \\
\hat{k}_d(u_1 - v_1) = \rho_2 \ddot{v}_1 \quad, \\
\hat{k}_d(u_2 - v_2) = \rho_2 \ddot{v}_2 \quad, \\
\hat{k}_\theta(\phi - \theta) = I_2 \ddot{\theta} \quad.
\end{cases}$$
(38)

The propagation of a harmonic plane wave travelling along axis $\mathbf{k} = \{k_1 \quad k_2\}^T$ with angular frequency $\omega$ and polarization vector $\hat{\mathbf{U}} = \{\hat{\mathbf{u}}^T \quad \hat{\phi} \quad \hat{\mathbf{v}}^T \quad \hat{\theta}\}^T = \{\hat{u}_1 \quad \hat{u}_2 \quad \hat{\phi} \quad \hat{v}_1 \quad \hat{v}_2 \quad \hat{\theta}\}^T$ is obtained by substituting the harmonic motion $\mathbf{U} = \hat{\mathbf{U}} \exp[i(\mathbf{k} \cdot \mathbf{x} - \omega t)]$ in equation (38) so obtaining the following system of six linear homogeneous equations:

$$\begin{bmatrix} \begin{bmatrix} (2\mu+\lambda)k_1^2 + \\ -2Ak_1 k_2 + \\ (\mu+k)k_2^2 + \\ +\hat{k}_d - \rho_1 \omega^2 \end{bmatrix} & \begin{bmatrix} Ak_1^2 + \\ +(\mu+\lambda-k)k_1 k_2 + \\ -Ak_2^2 \end{bmatrix} & 2i(Ak_1 - kk_2) & -\hat{k}_d & 0 & 0 \\ \begin{bmatrix} Ak_1^2 + \\ +(\mu+\lambda-k)k_1 k_2 + \\ -Ak_2^2 \end{bmatrix} & \begin{bmatrix} (\mu+k)k_1^2 + \\ +2Ak_1 k_2 + \\ +(2\mu+\lambda)k_2^2 + \\ +\hat{k}_d - \rho_1 \omega^2 \end{bmatrix} & 2i(kk_1 + Ak_2) & 0 & -\hat{k}_d & 0 \\ -2i(Ak_1 - kk_2) & -2i(kk_1 + Ak_2) & \begin{bmatrix} S(k_1^2 + k_2^2) + \\ +4k + \\ +\hat{k}_\theta - I_1 \omega^2 \end{bmatrix} & 0 & 0 & -\hat{k}_\theta \\ -\hat{k}_d & 0 & 0 & \begin{bmatrix} \hat{k}_d + \\ -\rho_2 \omega^2 \end{bmatrix} & 0 & 0 \\ 0 & -\hat{k}_d & 0 & 0 & \begin{bmatrix} \hat{k}_d + \\ -\rho_2 \omega^2 \end{bmatrix} & 0 \\ 0 & 0 & -\hat{k}_\theta & 0 & 0 & \begin{bmatrix} \hat{k}_\theta + \\ -I_2 \omega^2 \end{bmatrix} \end{bmatrix} \begin{Bmatrix} \hat{u}_1 \\ \hat{u}_2 \\ \hat{\phi} \\ \hat{v}_1 \\ \hat{v}_2 \\ \hat{\theta} \end{Bmatrix} = \mathbf{0} .$$

(39)

The solution of the eigenproblem (39) $\mathbf{C}_{Hom}(\mathbf{k}, \omega) \hat{\mathbf{U}} = \mathbf{0}$, with $\mathbf{C}_{Hom}$ hermitian matrix, provides six dispersion functions $\omega(k_1, k_2)$ defined in the plane $(k_1, k_2)$ by the irreducible Brillouin domain shown in figure 6(c). The accuracy obtained by the continuum formulation may be appreciated by the following property $\mathbf{C}_{Hom}(\mathbf{k}, \omega) = \mathbf{C}_{Lag}(\mathbf{k}, \omega) + O(|\mathbf{k}|^3)$, which is general for periodic lattices, but that does not hold if the positive defined stiffness $S^+$ is assumed in the formulation. In the long wavelength limit $\lambda \to \infty$, namely $|\mathbf{k}| \to 0$, the following angular frequencies are obtained:

$$\omega_{ac1,2} = 0$$

$$\omega_{opt3,4} = \sqrt{1 + \frac{\rho_2}{\rho_1}} \, \omega_d \qquad (40)$$

$$\omega_{opt5,6} = \frac{\sqrt{2}}{2} \sqrt{1 + \frac{I_2}{I_1} + \left(\frac{\omega_\phi}{\omega_\theta}\right)^2 \mp \sqrt{\left[1 + \frac{I_2}{I_1} + \left(\frac{\omega_\phi}{\omega_\theta}\right)^2\right]^2 - 4\left(\frac{\omega_\phi}{\omega_\theta}\right)^2}} \, \omega_\theta$$

being $\omega_d = \sqrt{\hat{k}/\rho_2}$ and $\omega_\theta = \sqrt{k_\theta/J_2} = \sqrt{\hat{k}_\theta/I_2}$ the translational and rotational frequency of the resonator, respectively, and $\omega_\phi = \sqrt{4k/I_1}$ the long wave-length frequency of the lattice with no resonator. In analogy to the discrete model, two acoustic branches are obtained, together with two couples of optical branches. The first couple of optical branches start from the critical point ($\left.\frac{d\omega}{d|\mathbf{k}|}\right|_{|\mathbf{k}|=0} = 0$) with frequency $\omega_{opt3,4}$, while two distinct optical branches depart from the critical points corresponding to the distinct frequencies.

*5.2 Tetrachiral lattice*

For the tetrachiral lattice, the elasticity tensors are obtained from (27) and (28) noting that $A_{cell} = a^2$, and the constitutive equation is written in matrix form as follows:

$$\begin{Bmatrix} \sigma_{11} \\ \sigma_{22} \\ \sigma_{12} \\ \sigma_{21} \\ m_1 \\ m_2 \end{Bmatrix} = \begin{bmatrix} 2\mu & 0 & 0 & B & 0 & 0 \\ 0 & 2\mu & -B & 0 & 0 & 0 \\ 0 & -B & \kappa & 0 & 0 & 0 \\ B & 0 & 0 & \kappa & 0 & 0 \\ 0 & 0 & 0 & 0 & S & 0 \\ 0 & 0 & 0 & 0 & 0 & S \end{bmatrix} \begin{Bmatrix} \gamma_{11} \\ \gamma_{22} \\ \gamma_{12} \\ \gamma_{21} \\ \chi_1 \\ \chi_2 \end{Bmatrix}, \qquad (41)$$

where the four elastic moduli are related to the lattice parameters as follows

$$\mu = \frac{E_s}{2}\left(\frac{t}{l}\right)\left[\cos^2\beta + \left(\frac{t}{l}\right)^2 \sin^2\beta\right],$$

$$\kappa = E_s\left(\frac{t}{l}\right)\left[\sin^2\beta + \left(\frac{t}{l}\right)^2 \cos^2\beta\right],$$

$$B = -\frac{E_s}{2}\left(\frac{t}{l}\right)\left[1 - \left(\frac{t}{l}\right)^2\right]\sin\beta\cos\beta,$$

$$S = -\frac{E_s a^2}{12}\left(\frac{t}{l}\right)\left\{3\left[\sin^2\beta + \left(\frac{t}{l}\right)^2 \cos^2\beta\right] - \left(\frac{t}{l}\right)^2\left(\frac{l}{a}\right)^2\right\},$$

(42)

where in case of ligaments tangent to the ring the first three moduli have been obtained in Bacigalupo and Gambarotta, 2014a. In case of first order approximation of the generalized displacement field the positive defined stiffness is obtained $S^+ = \frac{E_s a^2}{12}\left(\frac{t}{l}\right)\left\{3\left[\sin^2\beta + \left(\frac{t}{l}\right)^2 \cos^2\beta\right] + \left(\frac{t}{l}\right)^2\left(\frac{l}{a}\right)^2\right\}$. Similarly, to the hexachiral honeycomb, a coupling is obtained between the extensional strains and the asymmetric strains through the elastic modulus $B$ which is an odd function of the parameter of chirality $\beta$, while the other elastic moduli are even functions. In case of symmetric macro-strain fields, the resulting classical fourth order elasticity tensor has the elastic moduli of the tetragonal system. The free wave equation of motion takes the form:

$$\begin{cases} 2\mu u_{1,11} + k u_{1,22} + B u_{2,11} - B u_{2,22} - B\phi_{,1} + k\phi_{,2} + \hat{k}_d\left(v_1 - u_1\right) = \rho_1 \ddot{u}_1 \ , \\ B u_{1,11} - B u_{1,22} + k u_{2,11} + 2\mu u_{2,22} - k\phi_{,1} - B\phi_{,2} + \hat{k}_d\left(v_2 - u_2\right) = \rho_1 \ddot{u}_2 \ , \\ S\phi_{,11} + S\phi_{,22} + B u_{1,1} + B u_{2,2} - k u_{1,2} + k u_{2,1} - 2k\phi + \hat{k}_\theta\left(\theta - \phi\right) = I_1 \ddot{\phi} \ , \\ \hat{k}_d\left(u_1 - v_1\right) = \rho_2 \ddot{v}_1 \ , \\ \hat{k}_d\left(u_2 - v_2\right) = \rho_2 \ddot{v}_2 \ , \\ \hat{k}_\theta\left(\phi - \theta\right) = I_2 \ddot{\theta} \ . \end{cases}$$

(43)

The plane harmonic waves are obtained by solving the eigenvalue problem:

$$\begin{bmatrix} \begin{bmatrix} 2\mu k_1^2 + kk_2^2 + \\ \hat{k}_d - \rho_1\omega^2 \end{bmatrix} & B(k_1^2 - k_2^2) & -i(Bk_1 - kk_2) & -\hat{k}_d & 0 & 0 \\ B(k_1^2 - k_2^2) & \begin{bmatrix} kk_1^2 + 2\mu k_2^2 + \\ \hat{k}_d - \rho_1\omega^2 \end{bmatrix} & -i(kk_1 + Bk_2) & 0 & -\hat{k}_d & 0 \\ i(Bk_1 - kk_2) & i(kk_1 + Bk_2) & \begin{bmatrix} S(k_1^2 + k_2^2) + \\ +2k + +\hat{k}_\theta - I_1\omega^2 \end{bmatrix} & 0 & 0 & -\hat{k}_\theta \\ -\hat{k}_d & 0 & 0 & [\hat{k}_d - \rho_2\omega^2] & 0 & 0 \\ 0 & -\hat{k}_d & 0 & 0 & [\hat{k}_d - \rho_2\omega^2] & 0 \\ 0 & 0 & -\hat{k}_\theta & 0 & 0 & [\hat{k}_\theta - I_2\omega^2] \end{bmatrix} \begin{Bmatrix} \hat{u}_1 \\ \hat{u}_2 \\ \hat{\phi} \\ \hat{v}_1 \\ \hat{v}_2 \\ \hat{\theta} \end{Bmatrix} = \mathbf{0}$$

(44)

Six dispersion functions are obtained $\omega(k_1, k_2)$ defined in the plane $(k_1, k_2)$ by the irreducible Brillouin domain shown in figure 12(d). In the long wavelength limit $\lambda \to \infty$ the frequencies are those given by equation (40) where the long wave-length frequency of the lattice with no resonator is $\omega_\phi = \sqrt{2k/I_1}$.

## 6. Influence of the chirality and of the local resonators on the acoustic band structure

To investigate the sensitivity of the acoustic band structure and the formation of stop bands as a possible consequence of the chiral geometry of the microstructure and of the presence of local resonator, some example have been considered and analysed for both the hexachiral and the tetrachiral microstructure. For both the geometries, three cases have been considered which correspond, respectively, to the achiral geometry $(\beta = 0)$, to the intermediate chirality $(\beta = 10°)$ and to the maximum chirality $(\beta = \beta_m)$ that is attained when the ligaments are connected to the ring in the tangent point. The geometric parameters of the microstructure are those taken from Alderson *et al.*, 2010, and Bacigalupo and Gambarotta, 2014a, and are here considered in dimensionless form. The ratio between the mean radius of the ring to the cell size $r/a = 1/5$, the ratio between the ligament thickness and the cell size $t/a = 3/50$, the ratio $R/r = 1/2$ between the radius of the rigid mass of the resonator and the mean radius of the ring are considered, respectively. The following ratio defining the elastic modulus of the soft coating is assumed $E_a/E_s = 1/10$ with $\nu_a = 0.3$.

The acoustic band structures of both hexa- and tetra-chiral microstructures have been obtained in the first irreducible Brillouin zone (see Brillouin, 1953) though the discrete model (based of the Floquet-Bloch theory) formulated in Section 2. They are represented in the following in terms of the arch-length $\Xi$ in the dimensionless plane $(k_1 a, k_2 a)$ (see Figure 6(b) and 12(b)), and the dimensionless frequency $\omega a / \sqrt{E_s / \rho_s}$, being $E_s$ the Young modulus of the ligaments and $\rho_s$ the mass density of the ring. The arch-length $\Xi$ detects the distance between a generic point of the boundary of the first irreducible Brilluoin zone and the origin point $\mathbf{k} = \mathbf{0}$ (see Figure 6(b) and 12(b)).

Firstly, the band structure of the chiral lattice without internal resonators is described. In this case, the degrees of freedom for each ring are three and the band structure comprises three dispersion curves in the first irreducible Brillouin zone. The dispersion functions have two acoustic branches and an optical branch beginning at a critical point (in which the group velocity is zero, i.e. $v_g = 0$) with an dimensionless angular frequency that can be derived from equation (14), i.e.

$$\frac{\omega a}{\sqrt{\frac{E_s}{\rho_s}}} = \sqrt{\frac{n}{4\pi}\left(\frac{a}{l}\right)\left(\frac{a}{r}\right)^2 \left[\sin\beta + \left(\frac{t}{a}\right)^2 \left(\frac{a}{l}\right)^2 \cos\beta\right]} \qquad (45)$$

being $l/a = \cos\beta - \sqrt{(2r/a)^2 - \sin^2\beta}$, and $n$=4-6 for tetrachiral and hexachiral lattice, respectively. The angular frequency, which identifies the critical point and that influences the band structure of the beam-lattice increases with the angle of chirality $\beta$. Then, several cases having a different angle of chirality are examined in the presence of local resonators. Finally, some examples are considered with the aim to investigate the accuracy of the results provided by the equivalent continuum model, i.e. the micropolar generalized model formulated in Section 4. As this model is inherently formulated for the cases of moderately long waves, the comparison between the results of the discrete model and those of the equivalent continuum model is displayed in a homothetic sub-region of the irreducible Brillouin zone (see Figure 6(c) and 12(c)).

*6.1. Hexachiral beam-lattice*

In the case of microstructure without resonators the Floquet-Bloch spectrum is shown in Figure 6(a) in the first irreducible Brillouin zones shown in Figure 6(b) for three different values of the angle $\beta$. In these diagrams are visible both the acoustic branches and optical branch. In the case of achiral geometry ($\beta = 0$) is observed for low frequencies both a crossing point, between the optical branch and the second acoustic branch, and a veering phenomenon, namely to repulsion of the dispersion branches (see Phani *et al*., 2006), between the optical branch and the first acoustic branch. For each of the three considered values of $\beta$ a crossing of frequencies is observed for the dimensionless coordinate $\Xi_1 = 4/3\pi$. Finally, it is worth to note that the band structure of the three considered beam-lattices does not exhibit stop bands.

The plane acoustic behavior of the chiral lattice with internal resonators in case of maximum chirality ($\beta = \beta_m \simeq 23.6°$) is shown in the diagram of figure 7(a) where the first three dispersive curves are plotted for three different ratios $\rho_a/\rho_s$ between the mass density of the rigid mass of the resonator to the mass density of the rigid ring. It is worth to note that when $\rho_a/\rho_s \to 0$ and $E_a/E_s \to 0$, the Floquet-Bloch spectrum of the chiral lattice with resonator tends to that one of chiral lattice without resonator. In fact, the optical branches departing from the point with frequencies $\omega_{opt3,4}$ and $\omega_{opt6}$, are nearly independent of the wave number for small values of the ratios $\rho_a/\rho_s$ and identify the natural frequencies of the resonators. Moreover, an increasing of the ratio $\rho_a/\rho_s$ induces a decrease of the frequencies $\omega_{opt3,4}$ and $\omega_{opt5,6}$ from which the optical branches depart. As a consequence, the interaction of these branches is obtained with the acoustical ones and with the first optical branch. Such interaction between the branches of the spectrum can lead to the formation of frequency band-gaps. Indeed, it is noted that increasing the mass density of the resonator with respect to that one of the ring ($\rho_a/\rho_s = 10$), a band gap is obtained between the optical branch and the acoustical ones. This circumstance does not occur for $\rho_a/\rho_s = 1/10$ and $\rho_a/\rho_s = 1$ because the effect of the resonator is limited being the corresponding natural frequencies much higher than $\omega_{opt5}$. As a consequence, a limited interaction between the optical branches (departing from the frequencies $\omega_{opt3,4}$ and $\omega_{opt6}$) with the first optical branch and the acoustic branches appears. The full band structure for the case $\rho_a/\rho_s = 10$ shows the presence of a second band-gap at higher frequencies for $\omega \in (\omega_{opt5}, \omega_{opt3})$ between the third and the fourth optical branch.

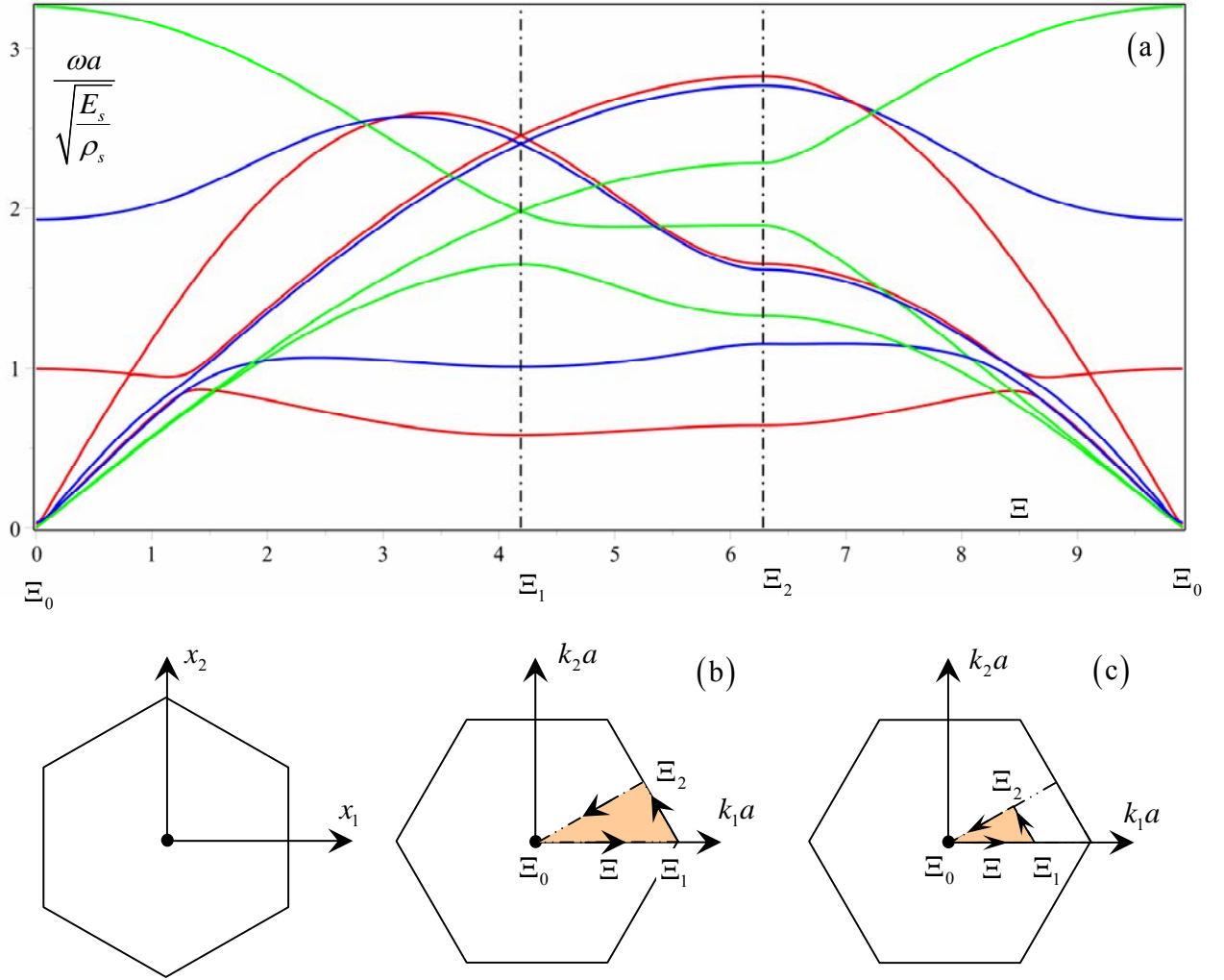

Figure 6. (a) Influence of the angle of chirality on the band structure of the hexachiral lattice without resonator in the irreducible Brillouin zone ($\beta = 0$ red; $\beta = 10°$ blue; $\beta = \beta_m \simeq 23.6°$ green). (b) Periodic cell and Brillouin zone (highlighted in orange the irreducible Brillouin zone); (c) Subdomain of the irreducible Brillouin zone.

Decreasing the chirality of the lattice ($\beta = 10°$), a band structure of the first three frequencies is obtained, which has no band gaps and which highlights various phenomena of crossing between the acoustical branches and the optical branch (see the diagrams in Figure 8(a)). Only in the case of high mass density of the resonator ($\rho_a / \rho_s = 10$) a band gap is obtained, which is shown in Figure 8(b) in the full band structure, located in between the first optical branch and the upper optical branches (associated to the vibration modes of the resonator) for $\omega \in (\omega_{opt5}, \omega_{opt3})$. This outcome appears to be in analogy to the higher band gap obtained for the case of maximum chirality (figure 8(b)) previously considered. Furthermore, compared to the previous case of maximum chirality, in this case the low frequency band gap comprised in between the acoustical branches and the first optical branches is missing.

In the hexagonal lattice ($\beta = 0$) it is observed that the frequency $\omega_{opt5}$, from which the first optical branch departs, decreases with the decrease of the angle $\beta$ (see Figures 9(a)). The frequency $\omega_{opt5}$ also depends on the ratio $\rho_a / \rho_s$ but so as the least relevant respect to angle $\beta$. In particular, this frequency tends to decrease with the increase of the ratio $\rho_a / \rho_s$. In addition, it may be noted that for all the mass density ratios $\rho_a / \rho_s$ considered, the phenomena of veering and crossing between the acoustical and the first optical branch is observed. The full band structure shown in figure 9(b) for $\rho_a / \rho_s = 10$ a wider band gap is obtained, though remaining unchanged the maximum frequency $\omega_{opt3}$ which is independent of $\beta$, while depends on the ratio $\rho_a / \rho_s$.

With the assumed values of the geometric parameters, it has been obtained that higher values of the angle of chirality imply the formation of two band gaps, the first one at low frequency (i.e. between the acoustical branches and the first branch optical). On the other side, in case of achiral lattice (i.e. of hexagonal lattice) only a band-gap with wider frequency range and located between the first optical branch and the higher optical branches is obtained. From the comparison of the band structures obtained in case of local resonators (see Figures 7(b), 8(b) and 9(b)) with those ones without resonators (see Figures 6(a)), shows the important effect of the last ones on the formation of the frequency band gaps. On the countrary, the effects of the chirality on the formation of the frequency band-gaps seem to be more limited.

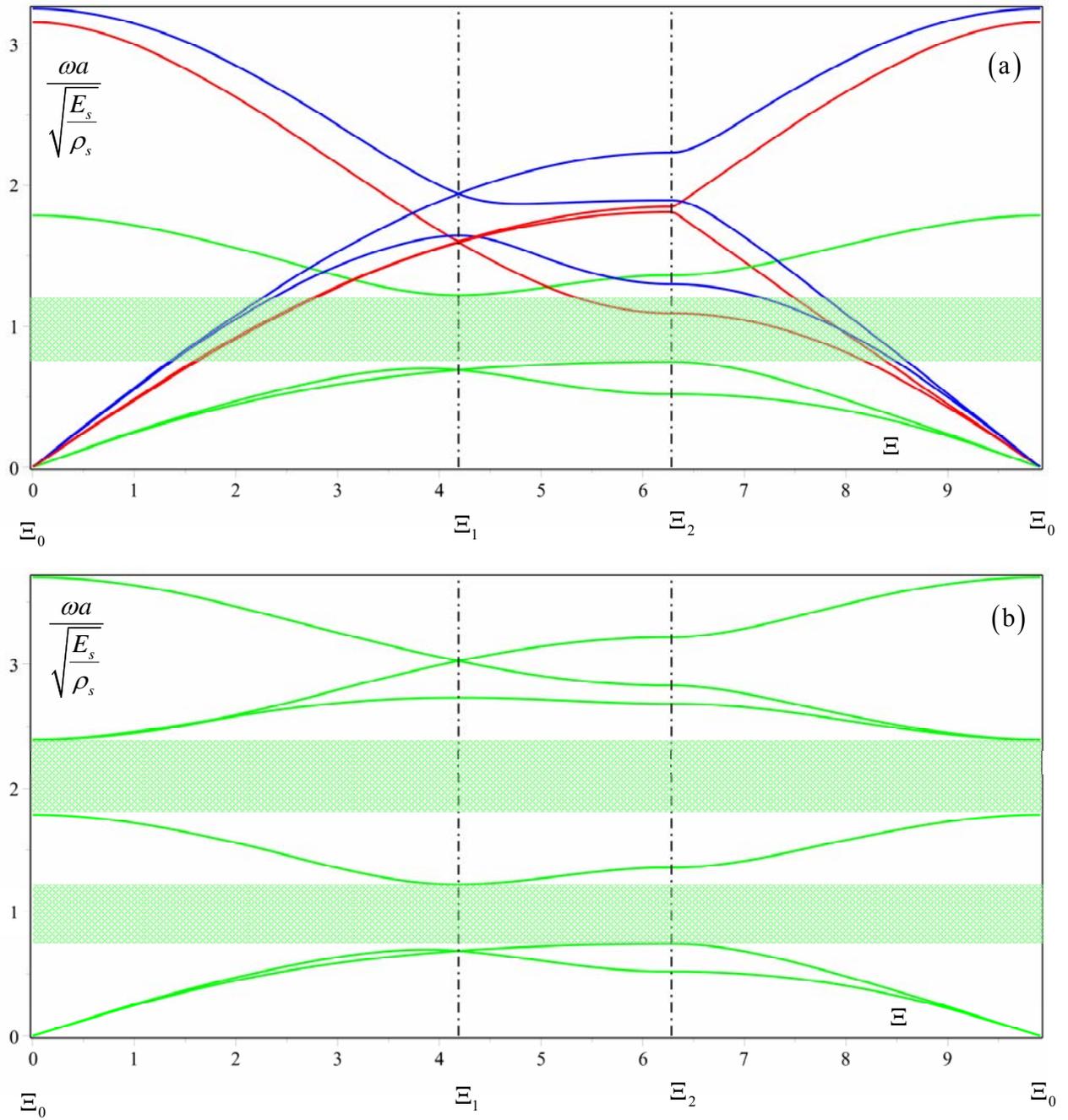

Figure 7. Band structure of hexachiral lattice with resonators $\beta = \beta_m \simeq 23.6°$: influence of the mass density ratio ($\rho_a/\rho_s = 1$ red; $\rho_a/\rho_s = 1/10$ blue; $\rho_a/\rho_s = 10$ green). (a) First three branches; (b) Full band structure ($\rho_a/\rho_s = 10$).

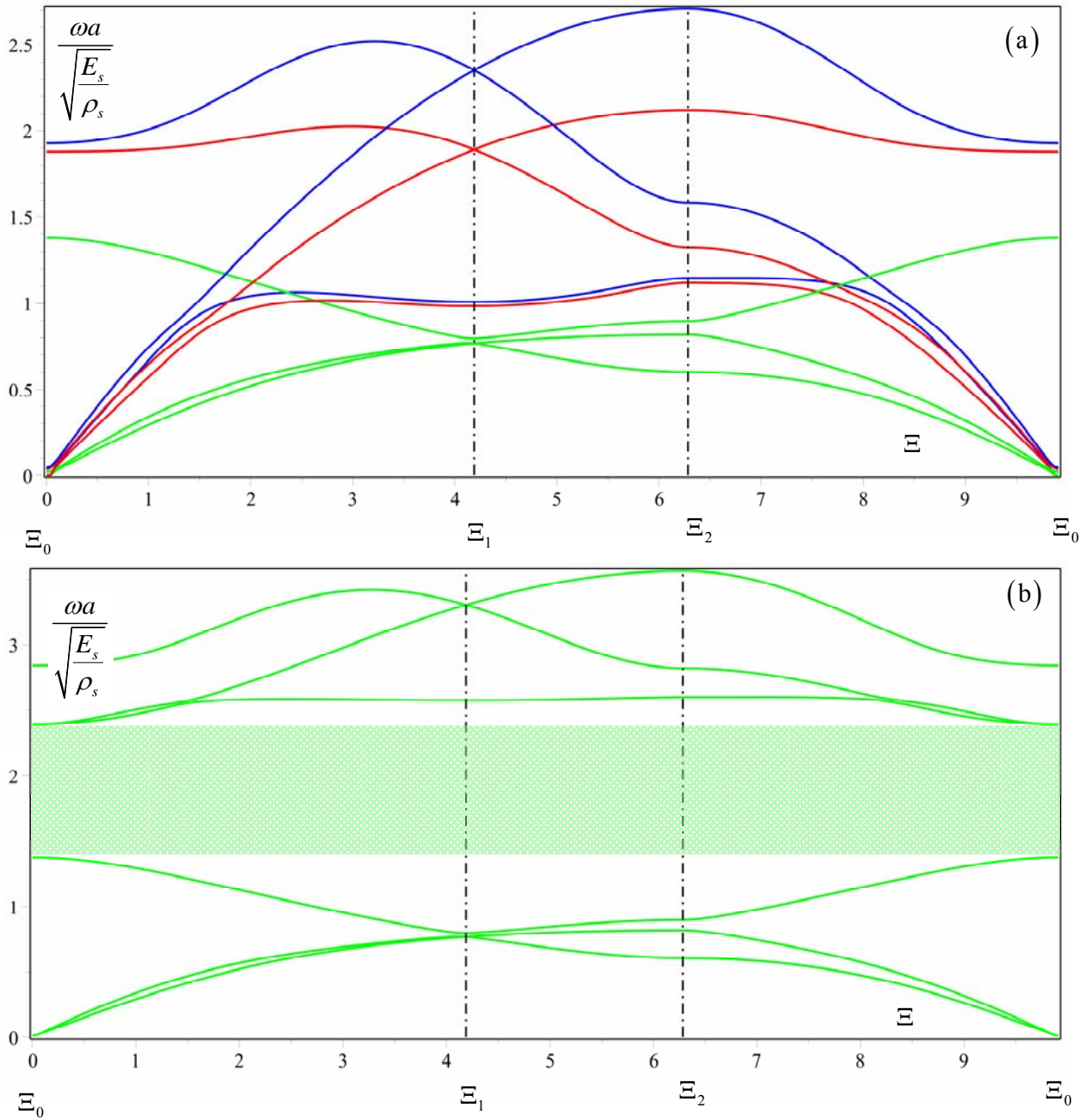

Figure 8. Band structure of hexachiral lattice with resonators $\beta = 10°$: influence of the mass density ratio ($\rho_a/\rho_s = 1$ red; $\rho_a/\rho_s = 1/10$ blue; $\rho_a/\rho_s = 10$ green). (a) First three branches; (b) Full band structure ($\rho_a/\rho_s = 10$).

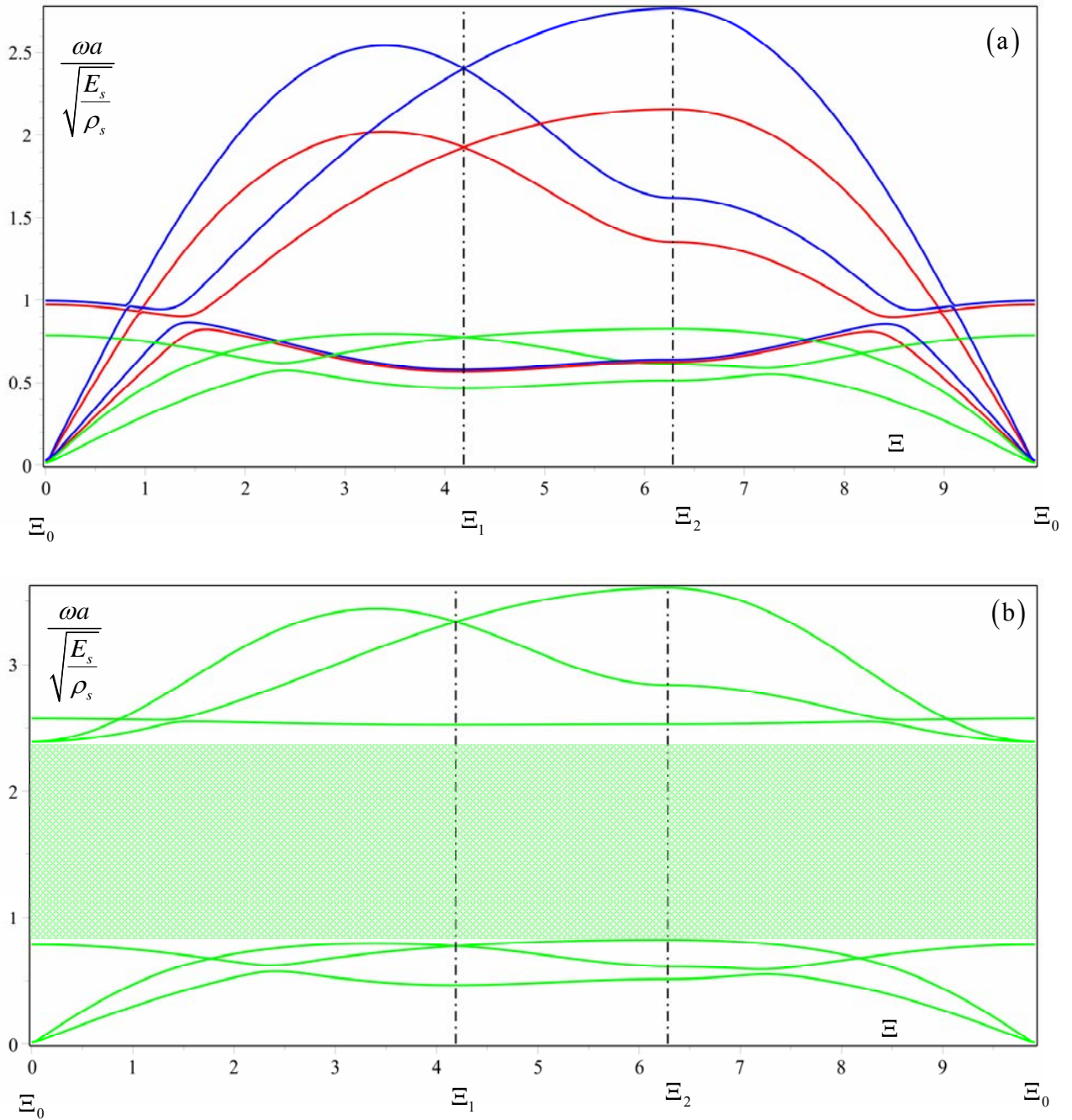

Figure 9. Band structure of hexagonal lattice with resonators $\beta = 0$: influence of the mass density ratio ($\rho_a / \rho_s = 1$ red; $\rho_a / \rho_s = 1/10$ blue; $\rho_a / \rho_s = 10$ green). (a) First three branches; (b) Full band structure ($\rho_a / \rho_s = 10$).

The accuracy of the generalized micropolar model derived in Section 5 is here analyzed for the case of maximum chirality $\beta = \beta_m \simeq 23.6°$. The band structure by the discrete model (for different values of the ratio $\rho_a/\rho_s$) is compared with the corresponding one from the continuum model, i.e. from the generalized micropolar model characterized by the overall elastic tensors $\mathbb{E}_s$ and $\mathbf{E}_s$ (with component $S$ of $\mathbf{E}_s$) (see equations (27), (28) and (36)). To this end, wave vectors in a homothetic sub-region of the first irreducible Brillouin zone are considered (see Figures 6(c)). The comparison is shown in the three dispersion diagrams in Figures 10(a), namely the two acoustic branches and the first optical branch. According to the property $\mathbf{C}_{Hom}(\mathbf{k},\omega) = \mathbf{C}_{Lag}(\mathbf{k},\omega) + O(|\mathbf{k}|^3)$ of the matrices of the discrete and continuum model, it follows that the accuracy of the generalized micropolar model seems to be acceptable with an error lesser than 10% for dimensionless wave numbers $|\mathbf{k}|a \leq 2/3\pi$, i.e. for wavelengths $\lambda \geq 3a$. Differences lower than 5% are obtained for $|\mathbf{k}|a \leq 4/9\pi$, i.e. for wavelengths $\lambda \geq 9/2\,a$.

The same type of diagrams are shown in Figure 10(b) for the case in which the generalized micropolar model is characterised by the elastic tensor $\mathbf{E}_s^+$ (with component $S^+$), see equations (27) and (29), obtained by assuming a first order approximation of the generalized displacement field. From this diagram it can be seen the low accuracy of the continuum model to represent the optical branches, while a better approximation is obtained for the acoustic branches.

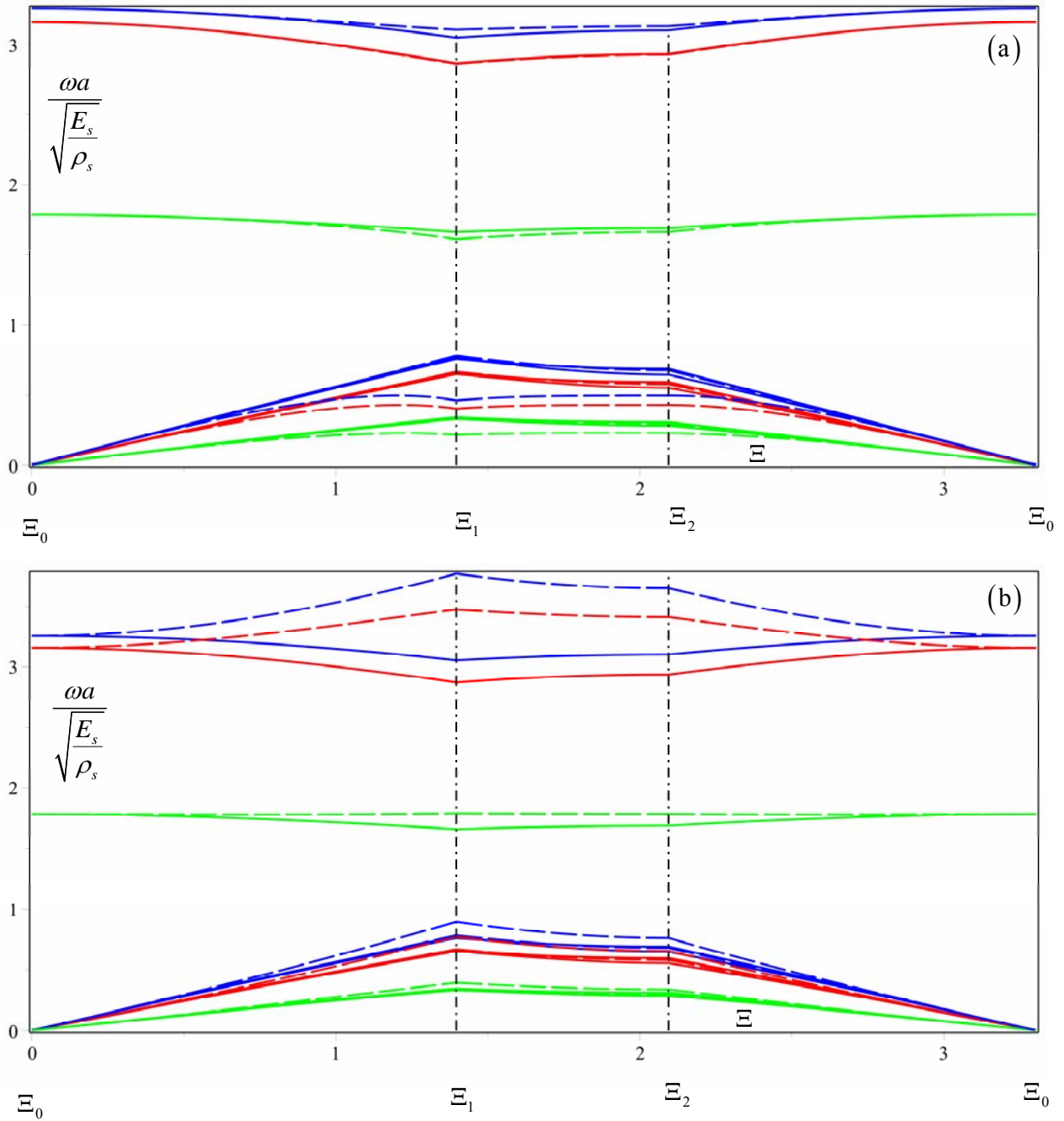

Figure 10. Dispersive functions for hexachiral lattice with resonator for $\beta \simeq 23.6°$. Comparison between the discrete model (continuous line) and the generalized micropolar continuum model (dashed line) of the first three branches in a subdomain of the irreducible Brillouin zone ($\rho_a/\rho_s = 1$ red; $\rho_a/\rho_s = 1/10$ blue; $\rho_a/\rho_s = 10$ green). (a) Constitutive constant $S$; (b) Constitutive constant $S^+$.

Finally, the comparison is extended in Figure 11 for propagation of harmonic waves along axis $x_1$ ($k_2 = 0$, $\Xi = k_1 a \in [0, 4/3\pi]$) for hexachiral lattice $\beta \simeq 23.6°$ (see Figure 11(a)) and for hexagonal (achiral) lattice $\beta = 0$ (see Figure 11(b)), where in black dashed line is shown the band structure of the generalized micropolar model.

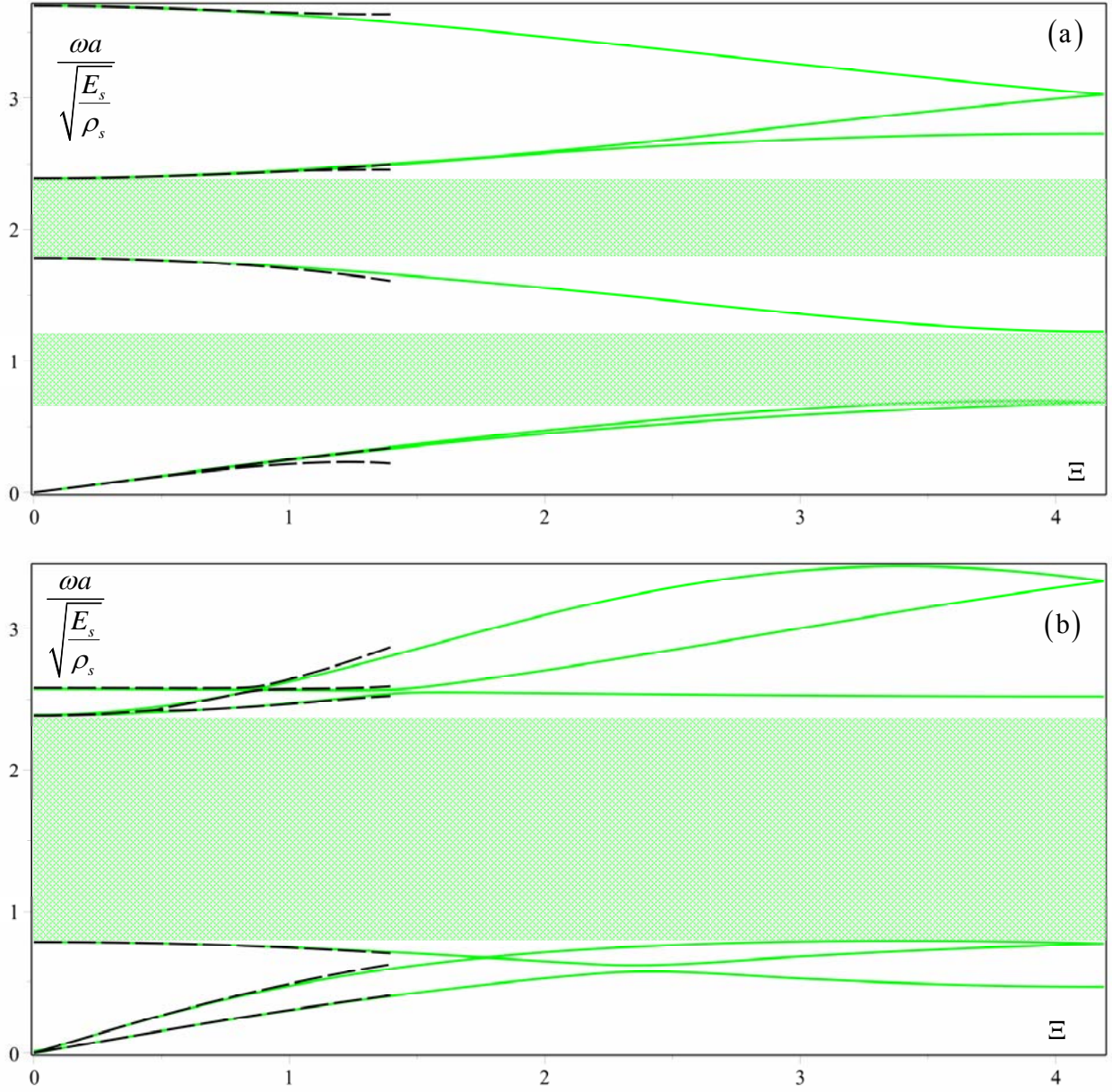

Figure 11. Dispersive functions and band gaps for wave propagation along $x_1$ axis ($\rho_a / \rho_s = 1/10$). Comparison between the discrete model (continuous line) and the generalized micropolar continuum model (dashed line). (a) $\beta = \beta_m \simeq 23.6°$, (b) $\beta = 0$.

*6.2. Tetrachiral beam-lattice*

The dispersion curves of the tetrachiral lattice without local resonators are shown for three different values of the angle β in the diagrams of Figure 12(a). In these diagrams, mapped in the first irreducible Brillouin zone shown in figure 12 (b), two acoustical and one optical branches may be identified. With respect to the band structure of the hexachiral lattice, the two acoustic branches appear here to be quite distinct. In particular, for $\Xi \in [0, \pi]$, the first acoustic branch is associated to the shear wave propagation along axis $x_1$, while the second acoustic branch is a pressure wave. For $\beta = \beta_m \simeq 23.6°$ the acoustic branch at higher frequency is very close to the optical branch in the range $\Xi \in [\pi, 2\pi]$. For the achiral tetragonal lattice the band structure appears to be similar to the first three branches obtained by Phani *et al.*, 2006. For all the considered cases, band gaps are not observed. Moreover, for $\Xi \in [0, \pi]$ and at low frequency a crossing point between the optical branch and the second acoustic branch is observed, while for $\Xi \in \left[2\pi, (2+\sqrt{2})\pi\right]$ a veering phenomenon between the optical branch and the second acoustic branch, i.e. a phenomenon of repulsion of the dispersion branches (see Phani *et al.*, 2006), takes place.

The acoustic behavior of the tetrachiral lattice with local resonators in case of maximum chirality ($\beta = \beta_m \simeq 23.6°$) is shown in the diagram of Figures 13(a), where the first three dispersion curves are plotted for three different ratios $\rho_a / \rho_s$. As for the hexachiral lattice, frequency band-gaps take place only for $\rho_a / \rho_s = 10$ (see Figures 13(b)). However, in this case a single band gap is obtained, in between the first optical branch and the upper one, in the range of frequencies $\omega \in (\omega_{opt5}, \omega_{opt3})$.

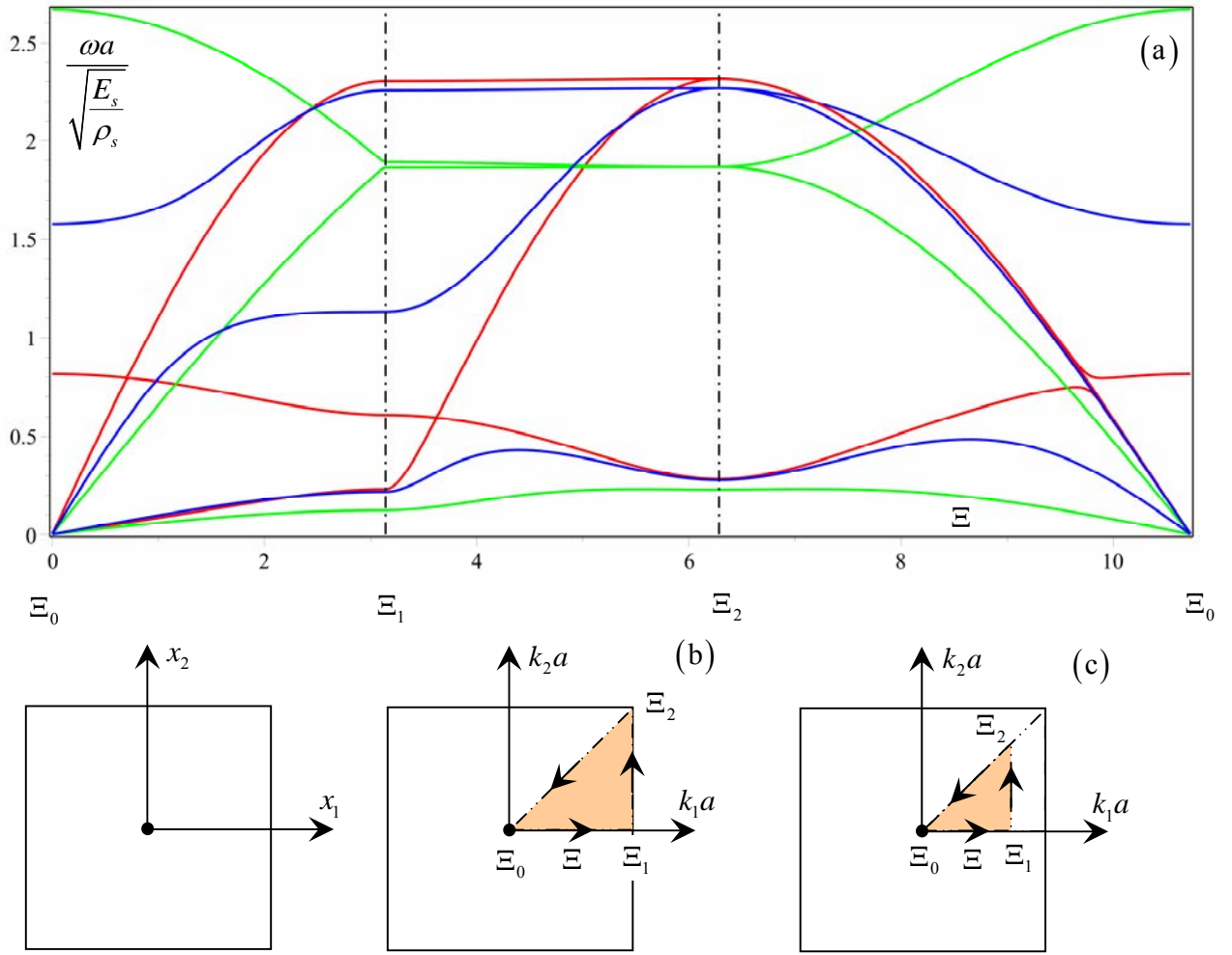

Figure 12. (a) Influence of the angle of chirality on the band structure of the tetrachiral lattice without resonator in the irreducible Brillouin zone ($\beta = 0$ red; $\beta = 10°$ blue; $\beta = \beta_m \simeq 23.6°$ green); (b) Periodic cell and Brillouin zone (highlighted in orange the irreducible Brillouin zone); (c) Subdomain of the irreducible Brillouin zone.

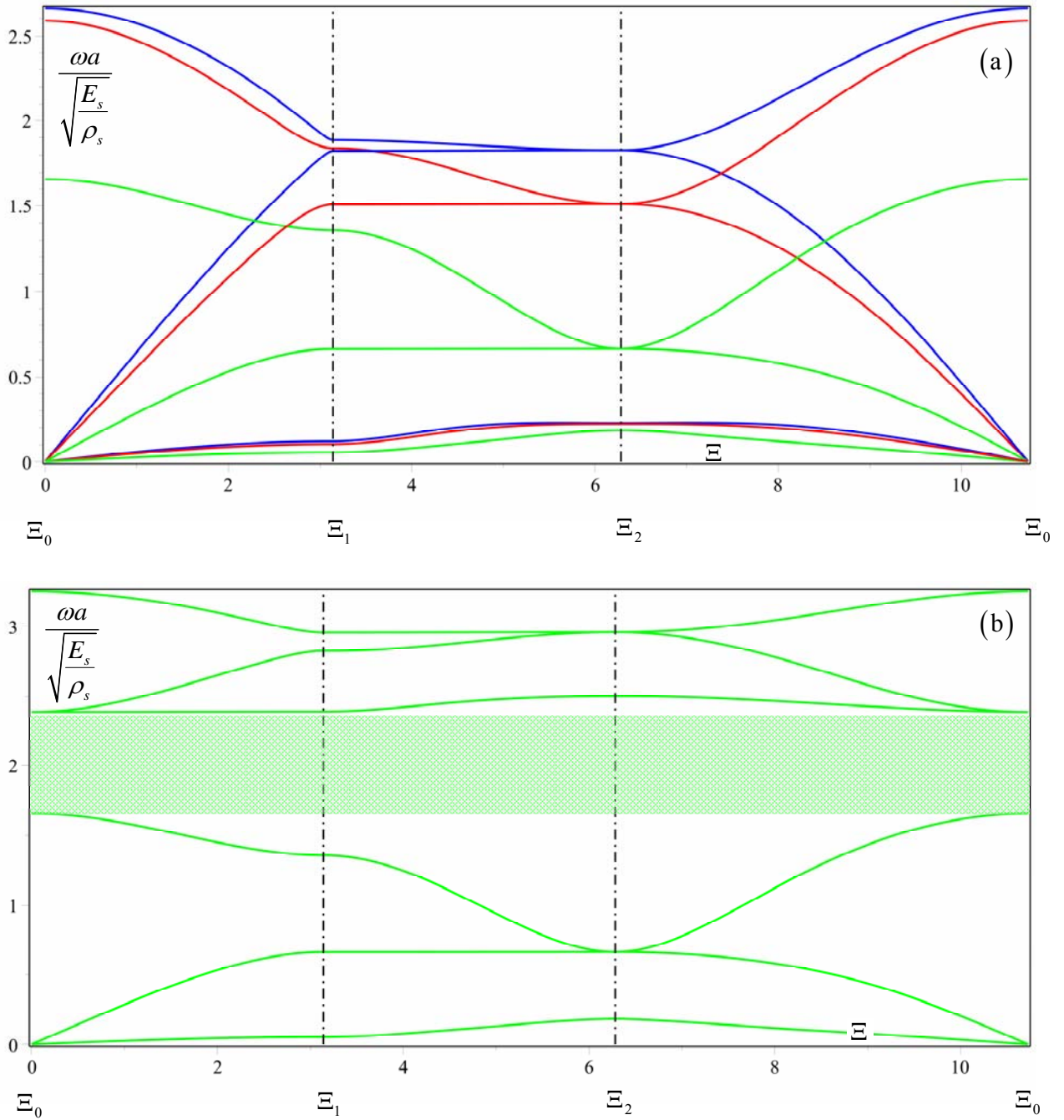

Figure 13. Band structure of tetrachiral chiral lattice with resonators $\beta = \beta_m \simeq 23.6°$: influence of the mass density ratio ($\rho_a/\rho_s = 1$ red; $\rho_a/\rho_s = 1/10$ blue; $\rho_a/\rho_s = 10$ green). (a) First three branches; (b) Full band structure ($\rho_a/\rho_s = 10$).

For lower values of the chirality angle ($\beta = 10°$), a similar band gap is obtained for $\rho_a / \rho_s = 10$ with a band structure which is similar to the previous one. Moreover, some critical points (with vanishing group velocity $v_g = 0$) at the vertices of the first irreducible Brillouin zones are identified. In particular, for $\Xi_2 = 2\pi$ the second acoustic branch and the first optical branch have the same angular frequency (see the diagrams in Figure 14(a)). From the diagrams of Figure 14(b) represented in the full Floquet Bloch spectrum, a band-gap is detects which appears to be wider than that shown in Figure 13(b) for $\beta \simeq 23.6°$. In fact, with decreasing angle $\beta$, the frequency $\omega_{opt5}$ decreases, while $\omega_{opt3}$ does not change. It is worth to note that the frequency $\omega_{opt5}$ depends on the ratio $\rho_a / \rho_s$ although in a more limited than the chirality $\beta$. In particular, the frequency $\omega_{opt5}$ tends to decrease with the increase of the ratio $\rho_a / \rho_s$.

In the tetragonal lattice ($\beta = 0$), the band structure exhibits a further decrease of the frequencies of the first optical branch (see Figures 15(a)). Consequently, the full band structure for $\rho_a / \rho_s = 10$, shown in figure 15(b), has a wider band-gap. For the values of the geometric parameters considered here, it is noted the presence of a single frequency band-gap regardless of the chirality $\beta$. However, the decrease of $\beta$ induces an increase of the band-gaps amplitude due to a decrease of the frequency $\omega_{opt5}$ from which the first optical branch departs. From the comparison of the band structures obtained in presence of local resonators (see Figures 13(b), 14(b) and 15(b)) with that of the chiral lattice without resonators (Figures 12(a)), it is remarked the important effect of the resonators on the formation of band-gaps in comparison to the effects of the chiral geometry of the lattice.

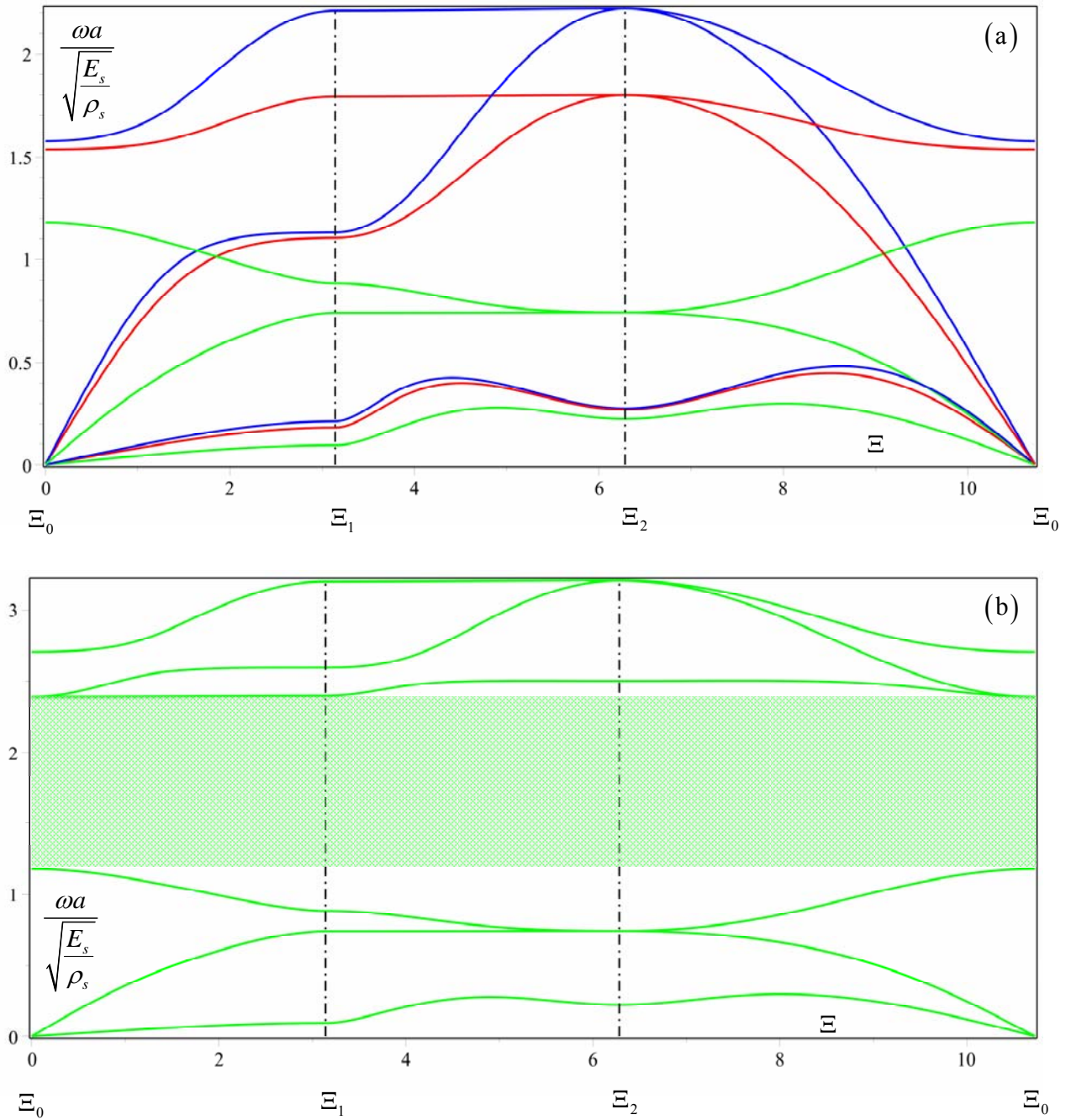

Figure 14. Band structure of tetrachiral lattice with resonators $\beta = 10°$: influence of the mass density ratio ($\rho_a/\rho_s = 1$ red; $\rho_a/\rho_s = 1/10$ blue; $\rho_a/\rho_s = 10$ green). (a) First three branches; (b) Full band structure ($\rho_a/\rho_s = 10$).

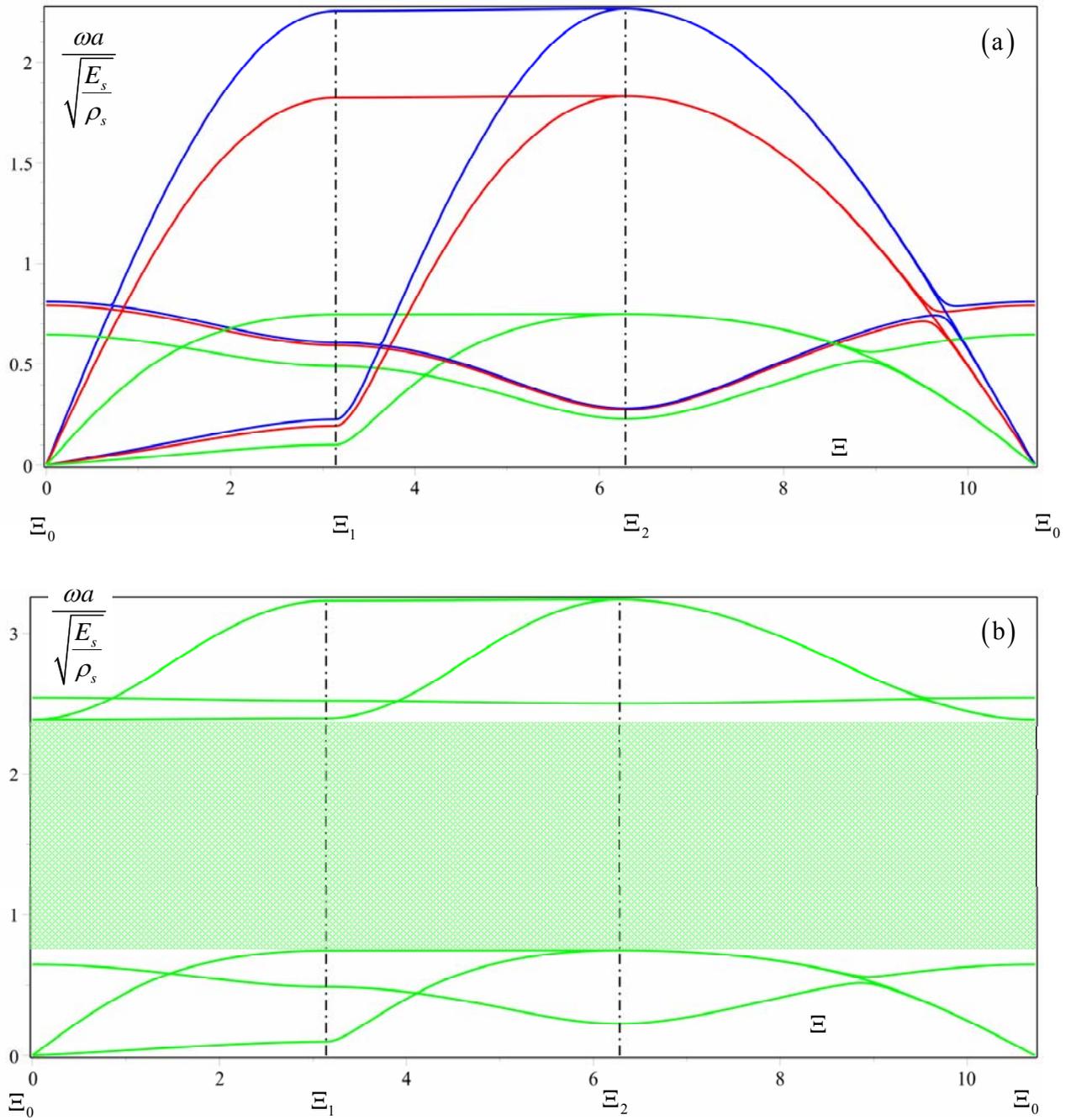

Figure 15. Band structure of tetragonal lattice with resonators $\beta = 0$: influence of the mass density ratio ($\rho_a / \rho_s = 1$ red; $\rho_a / \rho_s = 1/10$ blue; $\rho_a / \rho_s = 10$ green). (a) First three branches; (b) Full band structure ($\rho_a / \rho_s = 10$).

In the diagrams of Figure 16(a), the band structure by the discrete model for $\beta \simeq 23.6°$ and several values of the ratio $\rho_a / \rho_s$ is compared with that one by the generalized micropolar model, that depends on the overall elastic tensors $\mathbb{E}_s$ and $\mathbf{E}_s$ (with component $S$ of $\mathbf{E}_s$) (see equations (27), (28) and (42)). This comparison is carried out in the homothetic sub-domain of the first irreducible Brillouin zone (see Figures 12(c)) and is shown in Figures 16(a) for the two acoustic branches and for the first optical branch, in analogy to Figure 10 referred to the hexachiral lattice. It is worth to note that the band structure by the generalized micropolar model is in good agreement with that obtained by the discrete model; in particular, a better approximation is obtained on the optical branches. By taking, however, the second elastic tensor $\mathbf{E}_s^+$ (with component $S^+$) given by (29), the corresponding band structure is given in the diagrams in Figure 16(b), obtained by an approximation to the first order of the generalized displacement field. From this diagram it may be noted the low accuracy of such approximation in the optical branches, while the acoustic branches appear to be better approximated. In particular, the accuracy of the generalized micropolar model is within an error less than 10% for $|\mathbf{k}|a \leq 2/3\,\pi$, i.e. for wavelengths $\lambda \geq 3a$. Differences lower than 5% are obtained for $|\mathbf{k}|a \leq \sqrt{2}/3\,\pi$ i.e. $\lambda \geq 6/\sqrt{2}\,a$.

Finally, to get a better understanding of the accuracy of the continuum model, this comparison is shown in Figure 17 with reference wave propagation along the axis $x_1$, i.e. $\Xi \in [0, \pi/3]$, for tetrachiral lattice ($\beta \simeq 23.6°$), see Figure 17(a), and for tetragonal (achiral) lattice ($\beta = 0$), see Figure 17(b), where in black dashed line is shown the band structure by the generalized micropolar model.

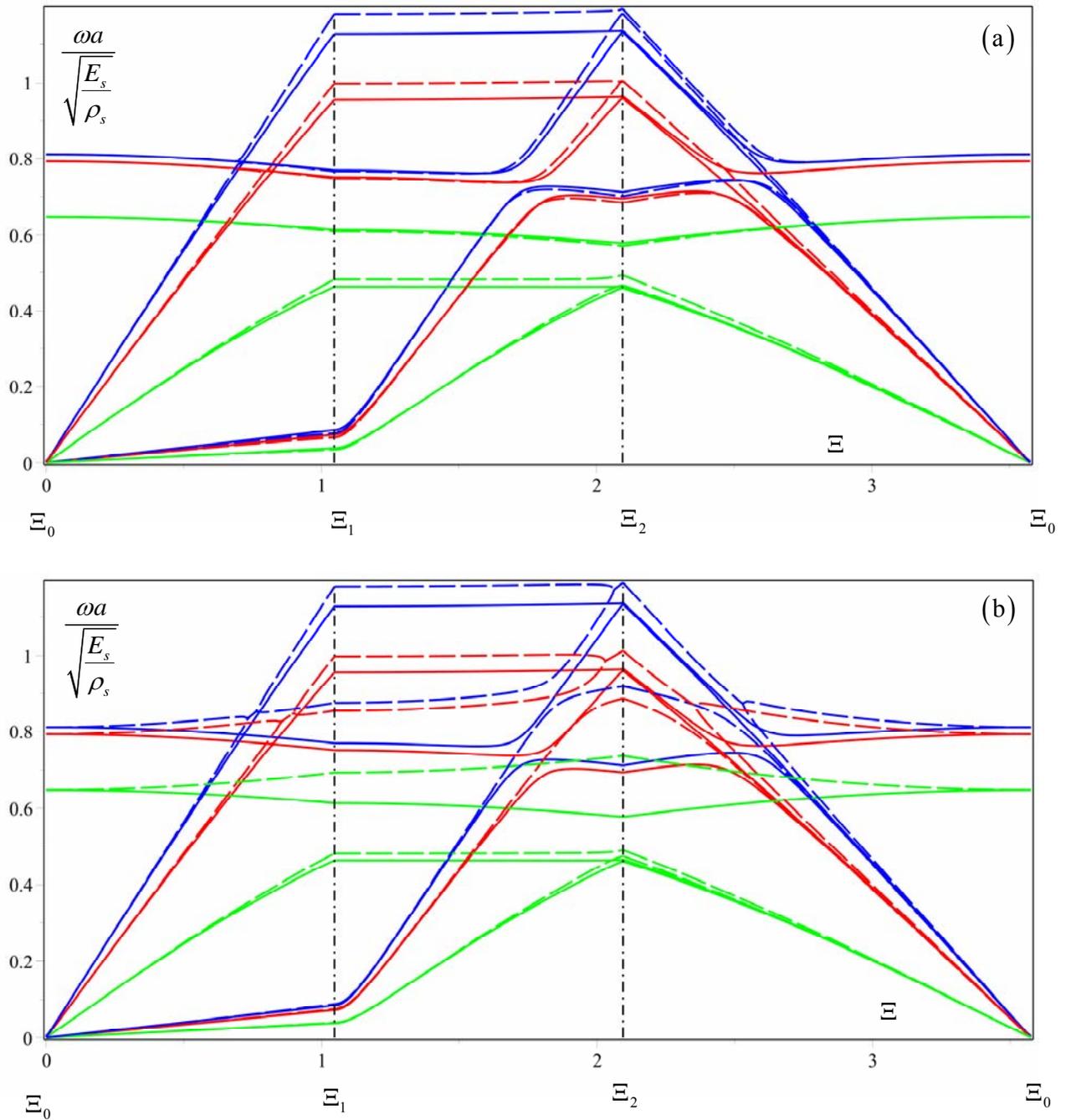

Figure 16. Dispersive functions for tetrachiral lattice with resonator for $\beta \simeq 23.6°$. Comparison between the discrete model (continuous line) and the generalized micropolar continuum model (dashed line) of the first three branches in a subdomain of the irreducible Brillouin zone ($\rho_a / \rho_s = 1$ red; $\rho_a / \rho_s = 1/10$ blue; $\rho_a / \rho_s = 10$ green). (a) Constitutive constant $S$; (b) Constitutive constant $S^+$.

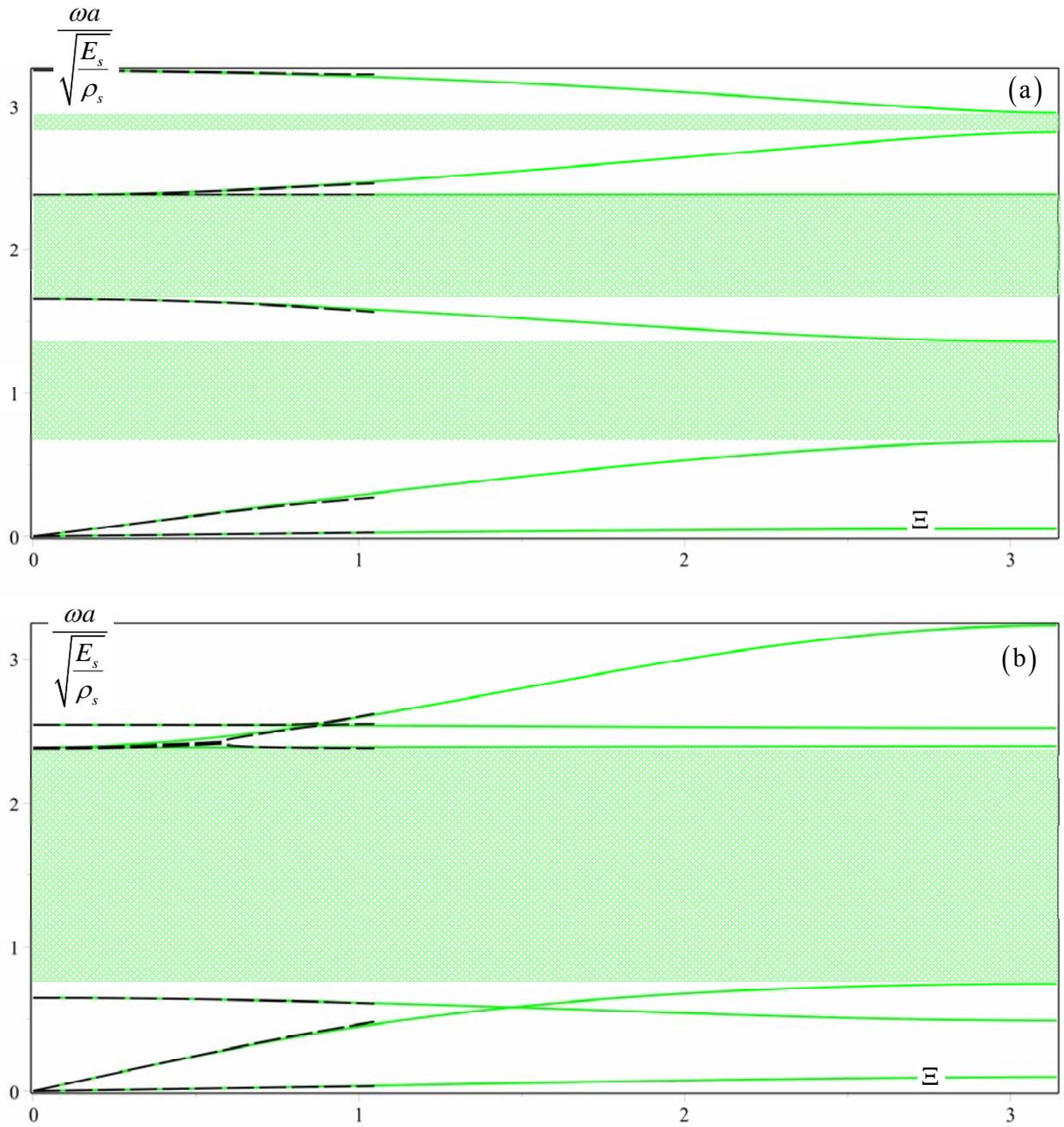

Figure 17. Dispersive functions and band gaps for wave propagation along $x_1$ axis ($\rho_a / \rho_s = 1/10$). Comparison between the discrete model (continuous line) and the generalized micropolar continuum model (dashed line). (a) $\beta \simeq 23.6°$ max chirality; (b) $\beta = 0$.

## 6. Conclusions

In this paper, a simplified model of periodic chiral beam-lattices containing local resonators has been formulated to obtain a better understanding of the influence of the chirality and of the dynamic characteristics of the local resonators on the acoustic behavior. In particular, the Floquet-Bloch spectrum and the occurrence of low frequency band-gaps are analysed through a discrete Lagrangian model. The simplified beam-lattices is made up of a periodic array of rigid heavy rings, each one connected to the others through elastic slender massless ligaments and containing an internal resonator made of a rigid disk in a soft elastic annulus. Two acoustic branches and two couples of optical branches characterize the band structure. The first couple branches off from a single critical point whose frequency $\omega_{opt3,4}$ depends on the ratio between the mass of the resonator and the mass of the ring and the translational frequency of the resonator. The branches of the second couple depart from two different critical points with frequency $\omega_{opt5,6}$ depending also on the rotational frequency of the resonator and on the frequency of the lattice with no resonators.

For prescribed geometry of the lattice, the dimensionless key parameters that control the acoustic behaviour and the wave propagation are: i) the ratio $\rho_a/\rho_s$, between the mass density of the rigid mass of the resonator to the mass density of the rigid ring; ii) the ratio $E_a/E_s$, between the elastic modulus of the soft coating of the resonator to the elastic modulus of the slender ligaments. The optical branches departing from the points with frequencies $\omega_{opt3,4}$ and $\omega_{opt6}$ are nearly independent of the wave number for small values of the ratios $\rho_a/\rho_s$ and identify the natural frequencies of the resonators. Moreover, an increasing of the ratio $\rho_a/\rho_s$ induces a decrease of the frequencies $\omega_{opt3,4}$ and $\omega_{opt5,6}$ from which the optical branches depart. Consequently, the interaction of these branches with the acoustical ones and with the first optical branch is obtained, which may lead to the formation of frequency band-gaps.

In hexachiral lattices, for high values of the chirality $\beta \approx \beta_m$ angle a low frequency band gap occurs between the first optical branch and the acoustical ones. A decrease of the chirality angle implies a progressively decrease of the optical branches starting from the frequencies $\omega_{opt5,6}$ with narrowing of this band gap. A higher band gap occurs between the first and the second optical branches departing from the critical point $\omega_{opt3,4}$, that is independent on the chirality angle. When decreasing the chirality angle, the lower bound of the higher band gap decreases and the band gap widens with a maximum amplitude for the achiral geometry. A similar behavior is observed for the tetrachiral geometry, with the difference that for high values of the angle of chirality only one band gap at high frequency is obtained, but not at low frequency (i.e. between the first optical branch and the acoustical branches). From the comparison between the band structures of lattices with internal resonators with those without resonators, within the simple model considered here, the important effect of the local resonators on the formation of the frequency band gaps is observed. On the contrary, the effects of the chirality seem to be more limited.

By approximating the generalized displacements of the rings of the discrete Lagrangian model as a continuum field and through an application of the generalized macro-homogeneity condition, a generalized micropolar equivalent continuum has been derived, together with the overall equation of motion and the constitutive equation given in closed form. The dispersive wave propagation in the equivalent continuum representative of both the hexachiral and the tetrachiral lattices has been obtained. The accuracy of the generalized micropolar model has been analyzed by comparing the band structures obtained by this model with those obtained using the Lagrangian discrete model. This comparison has been carried out with reference to a homotetic subdomain of the first irreducible Brillouin zone for harmonic acoustic waves with wave-lengths $\lambda \geq 3a$. A good approximation of the band structure of the discrete Lagrangian model is obtained by considering a second order approximation of the displacement field in the generalized micropolar model. Conversely, when considering a first-order approximation of the generalized displacement field, a good approximation of the acoustic branches is obtained with a lower approximation of optical branches.

**Appendix A – The translational and rotational stiffness of the local resonator**

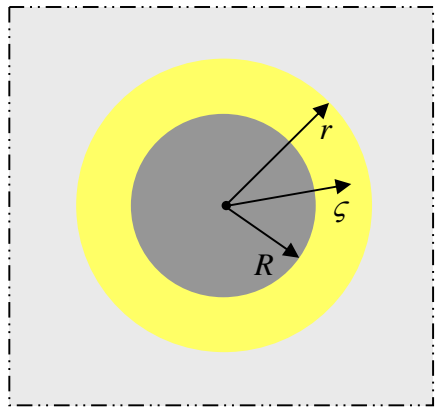

Figure A-1. Rigid disk contained in a soft elastic and isotropic annulus inserted into an external rigid body.

The translational and rotational stiffness of the resonator shown in Figure A-1 is here derived. Let us consider first the translation $u$ under plane stress conditions of the rigid disk having radius $R$ surrounded by an homogeneous, elastic, isotropic annulus with Young's modulus $E_a$ and Poisson's ratio $v_a$, having external radius $r$ (Figure A-1). The translational stiffness of the inner disk is evaluated through a FEM analysis by applying a distribution of forces with resultant $F$ to the internal disk. From the displacement $u$, coaxial with $F$, the stiffness $k_d = F/u$ is derived. In Figure A-2(a) the dimensionless translational stiffness $k_d/E_a$ as a function of the ratio $R/r$ is diagrammatically shown for different values of the Poisson's ratio $v_a$ having a weak effect on the translational stiffness $k_d/E_a$. As expected, a decrease of the annulus thickness (i.e. an increase of the geometric ratio $R/r$) induces a great increase of the stiffness.

The rotation $\theta$ of the rigid inner disk without translation $u$ is analysed by applying to the disk a distribution of forces having resultant torque $M$ and vanishing resultant force $F$. Because the symmetry of the system, the displacement $u(\varsigma)$ at a point of the annulus at a distance $\varsigma$ from the origin is tangent to the circumference of radius $\varsigma$ (see Figure A-1). Similarly, the shear stress $\tau(\varsigma)$ at the same point is tangent to the same circumference. Since the torque is $M = -2\pi\tau(\varsigma)\varsigma^2 = -2\pi\tau_R R^2$ by noting that $\tau(\varsigma) = \left(\dfrac{R}{\varsigma}\right)^2 \tau_R$, $\tau_R$ being the tangential stress on the internal boundary, and written the shear strain $\gamma = \dfrac{du}{d\varsigma} - \dfrac{u}{\varsigma} = \dfrac{\tau_R}{G_a}\left(\dfrac{R}{\varsigma}\right)^2$ in terms of $\tau_R$, with $G_a$ the shear elastic modulus of the annulus. By integration of the ODE $\dfrac{du}{d\varsigma} - \dfrac{u}{\varsigma} = \dfrac{\tau_R}{G_a}\left(\dfrac{R}{\varsigma}\right)^2$, with boundary condition $u(\varsigma = R) = \theta R$, the displacement is obtained in the form $u(\varsigma) = \theta\varsigma + \dfrac{\tau_R}{2G}\dfrac{\varsigma^2 - R^2}{\varsigma}$. The shear stress on the internal circumference is obtained by imposing the boundary condition $u(\varsigma = r) = 0$ and takes the form $\tau_R = 2G\dfrac{r^2}{R^2 - r^2}\theta$. Then, the resultant torque is obtained $M = 4\pi G\dfrac{R^2 r^2}{r^2 - R^2}\theta$ together with the rotational stiffness $k_\theta = 4\pi G\dfrac{R^2 r^2}{r^2 - R^2} = 4\pi G\left(\dfrac{\delta^2}{1 - \delta^2}\right)r^2$, with $\delta = R/r$. In Figure A-2(b) the dimensionless rotational stiffness $k_\theta/E_a r^2$ as a function of the ratio $R/r$ is diagrammatically shown for different values of the Poisson's ratio $v_a$.

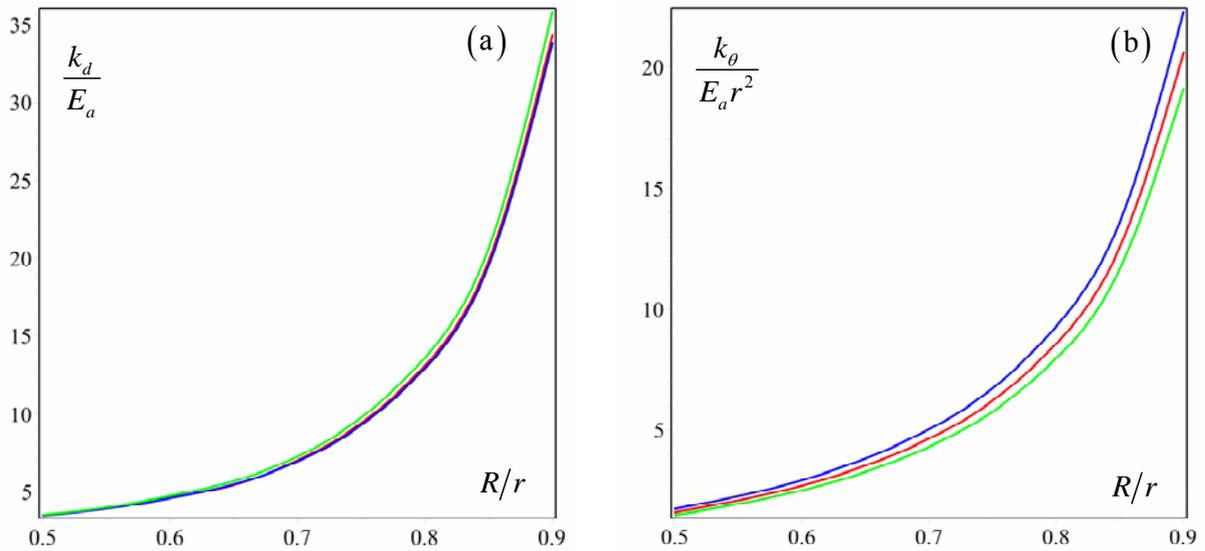

Figure A-2: (a) Dimensionless translational stiffness $k_d/E_a$ in terms of the ratio $R/r$; (b) Dimensionless rotational stiffness $k_\theta/E_a r^2$ of the ratio $R/r$. Influence of the Poisson ratio $v_a$: blue line $v_a = 0.2$; red line $v_a = 0.3$; green line $v_a = 0.4$.